\documentclass[number,review]{elsarticle}

\usepackage{lineno,hyperref}
\modulolinenumbers[5]

\journal{Elsevier Journal}

\usepackage[dvipsnames]{xcolor}
\definecolor{sPointColor}{HTML}{d300ff}
\definecolor{pPointColor}{HTML}{ff0081}
\usepackage{graphicx}
\usepackage{amsmath,bm,amsfonts,mathtools,amssymb}
\usepackage[toc,page]{appendix}
\usepackage{array}
\usepackage{paralist}
\usepackage{booktabs}
\usepackage{multirow}
\usepackage{multicol}
\usepackage{subcaption}
\usepackage{ulem}
\usepackage[outline]{contour}
\usepackage[T1]{fontenc}

\usepackage{pifont}
\newcommand{\cmark}{\ding{51}}%
\newcommand{\xmark}{\ding{55}}
\usepackage{diagbox}

\usepackage{listofitems}

\usepackage{orcidlink}

\usepackage{pgfplotstable}
\usepackage{pgfplots}
\pgfplotsset{compat=1.9}
\usetikzlibrary{plotmarks}
\usetikzlibrary{positioning}
\usetikzlibrary{matrix}
\usetikzlibrary{external}
\usetikzlibrary{calc}
\usetikzlibrary{arrows, arrows.meta}
\usetikzlibrary{decorations.markings}
\usetikzlibrary{shapes.geometric}
\usetikzlibrary{patterns}
\usepgfplotslibrary{fillbetween}

\pgfplotsset{select coords between index/.style 2 args={
    x filter/.code={
        \ifnum\coordindex<#1\fi
        \ifnum\coordindex>#2\fi
    }
}}

\usetikzlibrary{external}

\usepackage{algorithm}
\usepackage{algpseudocode}

\makeatletter

\newcommand{\norm}[1]{\left\lVert#1\right\rVert}

\newcommand{\mUnit}{\mathrm{M}}                                       
\newcommand{\tUnit}{\mathrm{T}}                                       
\newcommand{\vUnit}{\mathrm{L\,T^{-1}}}                               
\newcommand{\aUnit}{\mathrm{L\,T^{-2}}}                               
\newcommand{\lUnit}{\mathrm{L}}                                       
\newcommand{\ndUnit}{\mathrm{-}}                                      

\newcommand{\revs}[1]{\textcolor{black}{#1}}

\definecolor{schoolblack}{RGB}{35,31,32}
\definecolor{schoolred}{RGB}{240,78,35}
\definecolor{schoolgrey}{RGB}{110,110,110}
\definecolor{facblue}{RGB}{0,165,211}
\definecolor{mygrey}{RGB}{230,230,230}
\definecolor{streamblue}{HTML}{00c5bd}

\definecolor{cgreen}{RGB}{30,210,30}
\definecolor{xred}{RGB}{255,0,0}

\definecolor{cylgreen}{HTML}{00aa00}

\algnewcommand\algorithmicforeach{\textbf{for each}}
\algdef{S}[FOR]{ForEach}[1]{\algorithmicforeach\ #1\ \algorithmicdo}

\tikzset{cAlg/.style={ellipse,draw,minimum height=0.5cm,minimum width=0.8cm}}
\tikzset{cData/.style={ellipse,minimum height=0.5cm,minimum width=0.8cm,fill=gray,draw=gray,text=black,fill opacity=0.2,text opacity=1.0}}
\tikzset{cFunc/.style={rectangle,draw,minimum height=0.5cm,minimum width=0.8cm, align = center}}
\tikzset{cInfo/.style={rectangle,minimum height=0.5cm,minimum width=0.8cm,fill=gray,draw=gray,text=black,fill opacity=0.2,text opacity=1.0}}

\makeatother

\hypersetup{
    colorlinks=true,       
    linkcolor=black,          
    citecolor=black,        
    filecolor=black,      
    urlcolor=black,           
    allcolors=black
}

\begin{document}

\begin{frontmatter}

\title
{
    Extending the hybrid fictitious domain-immersed boundary method for Reynolds-averaged turbulence modeling
}

\address[ITCAS]
{
 Institute of Thermomechanics of the Czech Academy of Sciences,
 Dolej\v{s}kova 5, Prague 182~00, Czech Republic
}

\address[VSCHTDM]
{
 University of~Chemistry and Technology, Prague,
 Department of~Mathematics, Informatics and Cybernetics,
 Technick\'{a}~5, Prague 166~28, Czech Republic
}

\author[ITCAS,VSCHTDM]{Lucie Kub\'{i}\v{c}kov\'{a}\,\orcidlink{0000-0002-5047-8569}}
\author[ITCAS,VSCHTDM]{Martin Isoz\,\orcidlink{0000-0002-5862-2561}\corref{cor}}
\cortext[cor]{Corresponding author,
tel: \mbox{+420 26605 2832}.}
\ead{isozm@it.cas.cz}
\ead[url]{https://techmathgroup.isoz.eu}

\begin{abstract}
Engineering practice often calls for shape or topology optimization (TO) of fluid defining components, while the ever-increasing computing power allows the optimized cost functions to be based on computational fluid dynamics (CFD). However, a common bottleneck in CFD-based TO frameworks is the requirement for frequent remeshing. In order to alleviate this bottleneck, we propose an adaptation of an immersed boundary (IB) method variant, the hybrid fictitious domain-immersed boundary method, to leverage Reynolds-averaged Navier-Stokes (RANS) equations and wall functions. The main contribution of the present work lies in the design and open-source implementation of the IB-aware steady-state solution of the RANS equations via the SIMPLE algorithm in the OpenFOAM library. The proposed approach yields results consistent with the standard body-fitted CFD for the most common two-equation RANS models, Reynolds numbers from $10^1$ to $10^6$, and canonical benchmarks, including flow over a backward-facing step or an Ahmed body. Furthermore, given the intended application in TO, special emphasis is placed on the robustness and geometric generality, which is tested on a real-life engineering application in simulation of coolant flow in a die-casting die core.
\end{abstract}

\end{frontmatter}


\section{Introduction}
\label{sec:intro}
Immersed boundary methods (IBMs)~\citep{peskin1972} constitute a class of numerical techniques in computational fluid dynamics (CFD) that enable the simulation of flows around complex geometries without the need for body-fitted, geometry-conforming meshes. IBMs have been successfully used in a wide range of disciplines, including weather research~\citep{lundquist2010}, phase transition and multiphase models~\citep{blais2018, lavoie2022}, energy applications involving hydrofoils~\citep{lahooti2019, lim2022} and turbines~\citep{ouro2017, tafti2022}, as well as simulations of porous structures~\citep{das2018}, and particle-laden flows~\citep{breugem2012, saadat2018, isoz2022}.

In contrast to conventional body-fitted discretizations, immersed boundary methods represent the geometry via a scalar indicator field and a corresponding modification of the governing equations~\citep{mittal2023}. Consequently, geometric changes can be accommodated at significantly reduced computational costs, as regenerating the indicator field is substantially cheaper than rebuilding the mesh. This property makes IBMs particularly attractive for applications involving geometry optimization, as demonstrated, for example, in fluid–structure interaction studies~\citep{jenkins2016}. However, the application of IBMs to the geometry optimization of realistic engineering components requires their coupling with computationally efficient turbulence closures, in particular Reynolds-averaged Navier–Stokes (RANS) formulations relevant for high-Reynolds-number flows.

The coupling of IBMs with RANS remains challenging, with the primary difficulty being the treatment of near-wall regions~\citep{verzicco2023}. Due to the non-conforming nature of the underlying discretization, IBMs cannot leverage anisotropic mesh refinement in the wall-normal direction, which is essential for resolving steep velocity gradients in boundary layers within conventional approaches. As a result, IBMs often rely on wall modeling approaches. For example, in~\citet{kalitzin2002}, the velocity at the immersed boundary is reconstructed by linear interpolation and its tangential component is corrected by wall functions. A different two-layer approach was proposed by~\citet{capizzano2011}, where a thin-boundary layer equation is solved near the IB and RANS equations in the rest of the domain.

The inclusion of wall modeling approaches showed promising results, but in flows with high Reynolds numbers, spurious oscillations of pressure and wall shear stress occurred at the immersed boundary~\citep{constant2021}. \citet{cai2021} suggested avoiding such oscillations by interpolating the friction velocity rather than the raw velocity at the IB. Alternatively,~\citet{constant2024} proposed a pre-processing step to optimize the positions of interpolation points based on an estimate of the boundary layer thickness. Moreover,~\citet{troldborg2022} discussed another reason for instability arising from insufficient coupling between the momentum and turbulence equations. In particular, when the local velocity profile is used to calculate the friction velocity, which is then employed to correct the velocity at the immersed boundary, the feedback from the turbulence transport equations remains limited, thereby weakening the overall consistency of the coupled system.

A further difficulty arises when using IBMs for the solution of the incompressible RANS equations. As described in a review by~\citet{verzicco2023}, the forcing terms impose the correct velocity at the immersed boundary. However, the subsequent enforcement of the incompressibility constraint, i.e., the velocity field solenoidality, modifies the velocity through a global projection step. As a result, the velocity field is altered independently of the immersed-boundary forcing. To address this inconsistency, several remedies for the problem have been proposed, including additional iteration loops~\citep{fadlun2000}, Lagrange multipliers~\citep{kim2001}, or mass sources and sinks~\citep{uhlmann2005}.

Overall, no universally established framework exists for coupling immersed boundary methods with Reynolds-averaged turbulence models. Furthermore, many existing immersed-boundary approaches have been developed primarily for transient simulations, whereas geometry and topology optimization often benefit from computationally efficient steady-state solution strategies. Finally, only a limited subset of existing approaches is available as open-source implementations. This lack of standardization, combined with the methodological challenges discussed above, namely boundary-layer modeling and inconsistency with incompressibility enforcement, motivates the need for a robust and reproducible IBM–RANS formulation. 

In this work, we address this gap by presenting the openHFDIBRANS solver, a fully open-source implementation built within the OpenFOAM framework and designed to adhere to established CFD practices. The underlying immersed boundary formulation is based on the hybrid fictitious domain-immersed boundary (HFDIB) method introduced by~\citet{municchi2017} that was originally developed as a hybrid combining a direct forcing immersed boundary method for momentum and fictitious domain for the transport equations. In~\citep{municchi2017} and~\citep{municchi2018}, the method linkage to other similar methods is described and data about the method verification and validation are presented. Later, the HFDIB method was extended by~\citet{isoz2022, studenik2024} to incorporate the discrete element method (DEM) for arbitrarily shaped particles (openHFDIB-DEM). In this study, we extend the HFDIB formulation to RANS modeling (openHFDIBRANS), incorporating four standard two-equation turbulence models: the $k$–$\omega$, $k$–$\varepsilon$, $k$–$\omega$ SST, and realizable $k$–$\varepsilon$ models.

The proposed methodology combines ideas from existing approaches with tailored modifications aimed at improving robustness and consistency. In particular, wall modeling at the immersed boundary is based on wall-function formulations, where the friction velocity is evaluated from turbulence variables rather than reconstructed from local velocity profiles; this strategy enhances the coupling between momentum and turbulence equations. Furthermore, for high-Reynolds-number flows, the boundary velocity is not obtained via direct interpolation; it is instead corrected by the turbulent viscosity, which contributes to a tighter coupling of the transport equations. Finally, to alleviate inconsistencies between forcing and incompressibility enforcement, the divergence-free condition is imposed only in the fluid region, thereby preventing conflict with immersed-boundary forcing while still ensuring convergence in the steady-state limit. The performance and robustness of the solver are assessed across a representative set of benchmark problems, including the backward-facing step, flow over a smooth cylinder, and the Ahmed body, covering a range of flow regimes and geometric complexities. Lastly, the proposed methodology is applied on a real-life engineering problem focusing on a simulation of a coolant flow in a die-casting die core.

\section{Solver description}
\label{sec:method}
Here, the proposed variant of the immersed boundary (IB) method and its coupling with the Reynolds-averaged turbulence models are presented. The approach stems from the hybrid fictitious domain-immersed boundary (HFDIB) method, a direct forcing IB variant initially introduced by~\citet{municchi2017} and further refined by~\citet{isoz2022} for applications in particle-resolved direct numerical simulation of particle-laden flows; see~\citep{kotouc2024,studenik2024}. The original method is denoted as openHFDIB-DEM and is designed for direct numerical simulation with no inclusion of turbulence modeling, wall functions, or steady-state flow solution.

The extension of HFDIB for RANS presented in this work is denoted as openHFDIBRANS. A very early preliminary version of the method was described in our conference proceedings paper~\citet{kubickova2023}, but the current version is different on every level; from small details as inclusion of effective distance in the calculation of $y^+$; over tighter connection between transport equations; to usage of different formulation of continuity equation to enforce incompressibility primarily in fluid only. Overall, both the accuracy and convergence of the method were significantly improved.

The openHFDIBRANS solver is implemented using the open-source C++ library OpenFOAM~\citep{oF}. Its complete source code and examples are available from \href{https://github.com/techMathGroup/openHFDIBRANS}{https://github.com/techMathGroup/openHFDIBRANS}.

In the following, we first introduce the general governing equations. Afterwards, details on the IB treatment are provided. Particular emphasis is placed on distinguishing between resolved and unresolved boundary layer scenarios and evaluating wall functions at the immersed boundary.

\subsection{Governing equations}
\label{sub:equations}
The governing equations utilized in openHFDIBRANS are the steady-state Reynolds-averaged Navier-Stokes (RANS) equations complemented by two-equation turbulence closure models. The implemented form of the RANS equations is
\begin{equation}
\label{eq:hfdibrans}
\begin{array}{*1{>{\displaystyle}c}}
    \nabla \cdot (\bm{u} \otimes \bm{u}) = \nabla\cdot\left\{ \left(\nu + \nu_\mathrm{t}\right) \left[ \nabla\bm{u} + \nabla \bm{u}^{\text{T}} \right] \right\} - \nabla \tilde{p} + \bm{f}_\mathrm{ib}\,, \\ [0.3cm]
    \nabla \cdot \left[(1 - \lambda)\,\bm{u}\right] = 0\,,
\end{array}
\end{equation}
where $\bm{u}$ is the average velocity, $\nu$ the kinematic viscosity, $\nu_\mathrm{t}$ is the turbulent viscosity, and $\tilde{p}$ the average kinematic pressure. The effect of the immersed boundary is included in the term $\bm{f}_{\mathrm{ib}}$ and multiplicative factor $(1 - \lambda)$. In particular, the immersed boundary-induced source term $\bm{f}_\mathrm{ib}$ simulates the effect of the solid body on the flow momentum. Second, in the multiplicative factor, $\lambda$ is a phase indicator field. \revs{By taking the divergence of $\left[(1 - \lambda)\,\bm{u}\right]$, the continuity condition is effectively enforced only in the fluid and partially at the solid-fluid interface. To illustrate this, the divergence can be rewritten using the finite volume discretization as a sum of fluxes over the faces of a cell. With $\{\mathrm{f}\}_P$ being a set of faces of a cell $\Omega_P$ of volume $V_P$, the continuity condition in $\Omega_P$ becomes}
\begin{equation}
\label{eq:lambdaphi}
    \revs{
    \left[\nabla \cdot \left[(1 - \lambda)\,\bm{u}\right]\right]_P = \left[\nabla \cdot \left( \bm{u} - \lambda\,\bm{u}\right)\right]_P \approx \frac{1}{V_P} \sum_{\{\mathrm{f}\}_P}\left(\bm{u}_\mathrm{f} - \lambda_\mathrm{f}\,\bm{u}_\mathrm{f}\right)\cdot\bm{n}_\mathrm{f} A_\mathrm{f}\,,
    }
\end{equation}
\revs{where $\bm{n}_\mathrm{f}$ is the outer unit normal of the face and $A_\mathrm{f}$ is the face area. The subtraction of $\lambda_\mathrm{f}\,\bm{u}_\mathrm{f}$ suppresses spurious flux contributions originating inside the solid phase ($\lambda_\mathrm{f} = 1$) even though the velocity inside the solid should be uniform. At the interface, the subtracted term only relaxes the effect of the fluxes, but when convergence is reached, both the fluid ($\lambda_\mathrm{f} = 0$) and the solid have solenoidal velocity fields, and with a one-cell-wide interface, the continuity condition is consequently satisfied also in the relaxed interface cells ($0 < \lambda_\mathrm{f} < 1$).}

To complement the RANS equations, four turbulence closure models are available in openHFDIBRANS. The implemented models are the $k$-$\varepsilon$ model by~\citet{launder1974} and~\citet{tahry1983}, the $k$-$\omega$ model by~\citet{wilcox2006}, the Menter's $k$-$\omega$ shear stress transport (SST) model~\citep{menter1992}, and the realizable $k$-$\varepsilon$ model by~\citet{shih1995}.

To illustrate the turbulence model implementation, the $k$-$\omega$ model was chosen. Its equations~\citep{wilcox2006} are included in the following form
\begin{equation}
\label{eq:hfdibkomega}
\begin{array}{*1{>{\displaystyle}c}}
    \nabla \cdot (\bm{u}\, k) = \nabla \cdot \left[ \left(\nu + \frac{\nu_\mathrm{t}}{\sigma_{k_1}} \right) \nabla k \right] + P_k - \beta^*\, k\, \omega + S_\mathrm{ib}\,,\\[0.5cm]
    \nabla \cdot (\bm{u}\, \omega) = \nabla \cdot \left[ \left( \nu + \frac{\nu_\mathrm{t}}{\sigma_{\omega_1}} \right) \nabla \omega \right] + C_{\alpha 1}\, \frac{\omega}{k}\, P_k - C_{\beta 1}\, \omega^2 \,,\\[0.5cm]
\end{array}
\end{equation}
where $k$ is the turbulence kinetic energy, $\omega$ the specific dissipation of $k$, $P_k$ the production of $k$, and $S_\mathrm{ib}$ is another immersed boundary-induced source term. The remaining symbols represent model constants~\citep{wilcox2006}. For other turbulence models, both the $k$ and $\omega$ or $\varepsilon$ equations are included in the same manner. The $k$ equation has an additional source term $S_\mathrm{ib}$, and $\omega$ or $\varepsilon$ equations are the same as in the original formulations.

The steady-state governing equations~\eqref{eq:hfdibrans} and~\eqref{eq:hfdibkomega} are solved on $\Omega \subset \mathbb{R}^3$, an open, bounded, and connected set representing the spatial computational domain. Since the immersed boundary approach is of interest here, $\Omega$ is assumed to comprise both the solid phase $\Omega_\mathrm{s}$ and the fluid phase $\Omega_\mathrm{f}$.

\subsection{Immersed boundary-induced source terms}
\label{sub:sources}
In the governing equations, two immersed boundary-induced source terms are present: $\bm{f}_\mathrm{ib}$ in the momentum equation~\eqref{eq:hfdibrans}$_1$ and $S_\mathrm{ib}$ in the conservation equation of $k$~\eqref{eq:hfdibkomega}$_1$. These are computed as
\begin{equation}
\label{eq:qibalpha}
    \bm{f}_{\mathrm{ib}} = \alpha_{\bm{u}}\,\left[\,\mathcal{M}(\bm{u}_{\mathrm{ib}}) + \nabla \tilde{p}\,\right],\quad S_\mathrm{ib} = \alpha_k\ \mathcal{N}(k_\mathrm{ib})\,,
\end{equation}
where $\mathcal{M}$ and $\mathcal{N}$ are operators that stem from the governing equations as
\begin{equation}
\label{eq:operators}
\begin{array}{*1{>{\displaystyle}c}}
    \mathcal{M} (\bm{u}) = \nabla \cdot (\bm{u} \otimes \bm{u}) - \nabla \cdot\left\{ (\nu + \nu_\mathrm{t}) \left[ \nabla\bm{u} + \nabla \bm{u}^{\text{T}} \right] \right\}\,, \\ [0.3cm]
    \mathcal{N}(k) = \nabla \cdot (\bm{u}\, k) - \nabla \cdot \left[ \left(\nu + \frac{\nu_\mathrm{t}}{\sigma_{k_1}} \right) \nabla k \right] - P_k+ \beta^*\, k\, \omega\,.
\end{array}
\end{equation}
Moreover, the IB-induced source terms are constructed from 
\begin{inparaenum}[(i)]
    \item{mask fields $\alpha_{\bm{u}}$ and $\alpha_k$, which restrict the support of the source terms to the relevant regions of $\Omega$, and}
    \item{imposed fields $\bm{u}_\mathrm{ib}$ and $k_\mathrm{ib}$, defined so as to satisfy the required boundary conditions at the immersed boundary.}
\end{inparaenum}
In cells where the IB-induced source terms are active, the values of the imposed fields are enforced during an iterative solution process~\cite{municchi2017}. In the following, we first address the construction of the mask fields; subsequently, we consider the construction of the imposed fields and the overall solution algorithm, with an intermezzo devoted to the computation of $y^+$ at the immersed boundary.

\subsection{Mask fields construction}
\label{sub:lambda}
\begin{figure}[htbp]
    \centering
    \includegraphics{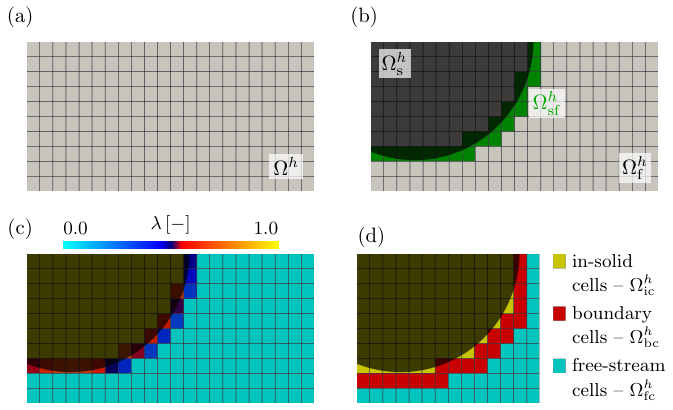}
    \caption{(a) Discrete spatial domain $\Omega^h$. (b) Overlay of a solid body over the domain from (a) with highlighted subdomains from~\eqref{eq:lamomega}. (c) Resulting $\lambda$ field. (d) Cell groups constructed based on the $\lambda$ field.}
    \label{fig:lamomega}
\end{figure}
To construct the mask fields $\alpha_{\bm{u}}$ and $\alpha_k$, the solid bodies need to be projected on the spatial domain $\Omega$. For the description, we will focus solely on discrete settings. Thus, let $\Omega^h\approx\Omega$ be a discrete spatial domain, see Figure~\ref{fig:lamomega}a. Based on the geometry of the solid body, $\Omega^h$ can be divided into three subdomains
\begin{equation}
\label{eq:lamomega}
    \Omega^h = \Omega^h_\mathrm{f} \cup \Omega^h_\mathrm{sf} \cup \Omega^h_\mathrm{s}\,,
\end{equation}
where $\Omega^h_\mathrm{f}$ contains cells that are fully immersed in fluid, $\Omega^h_\mathrm{s}$ contains those fully immersed in solid, and $\Omega^h_\mathrm{sf}$ those intersected by the fluid-solid interface, see Figure~\ref{fig:lamomega}b. The domain division is represented by an indicator scalar field $\lambda$, see Figure~\ref{fig:lamomega}c. With $\Omega^h_P\in\Omega^h$ being a computational cell, the $\lambda$ field is defined as
\begin{equation}
\label{eq:lambda}
    \lambda = \left\{
    \begin{array}{rc}
        0 & \text{if}\  \Omega^h_P \in \Omega^h_\mathrm{f} \\ [0.2cm]
        \tilde{\lambda} \in (0,1) & \text{if}\  \Omega^h_P \in \Omega^h_\mathrm{sf} \\ [0.2cm]
        1 & \text{if}\  \Omega^h_P \in \Omega^h_\mathrm{s}
    \end{array}
    \right.,\quad \tilde{\lambda} = \frac{1}{2}\left[1 - \text{tanh}\left( \frac{\sigma_\perp}{V(\Omega^h_P)^{\frac{1}{3}}} \right) \right]\,,
\end{equation}
where $V(\Omega^h_P)$ is the volume of the cell and $\sigma_\perp$ is the signed perpendicular distance from the cell center to the solid surface. The distance $\sigma_\perp$ is positive when the cell center is in the fluid and negative when in the solid. When the cell center is exactly at the fluid-solid interface, $\sigma_\perp$ is zero.

In openHFDIBRANS, the $\lambda$ field is used to classify cells into three groups. The groups are visualized in Figure~\ref{fig:lamomega}d. Each cell is classified as one of the following:
\begin{inparaenum}[(i)]
    \item{an in-solid cell ($\Omega^h_\mathrm{ic}$) if $\lambda \geq 0.5$,}
    \item{a boundary cell ($\Omega^h_\mathrm{bc}$) if $\lambda \in (0,0.5)$, or $\lambda = 0$ and there is an in-solid cell as the cell neighbor, or}
    \item{a free-stream cell ($\Omega^h_\mathrm{fc}$) if $\lambda = 0$ and none of the cell neighbors are in-solid cells.}
\end{inparaenum}

Based on the cell groups, the mask fields that restrict the support of the source terms~\eqref{eq:qibalpha} are defined as
\begin{equation}
\label{eq:alphauk}
    \alpha_{\bm{u}} = \alpha_k = \left\{
    \begin{array}{rl}
        1 & \text{if}\ \Omega^h_P \in \Omega^h_\mathrm{ic} \\ [0.2cm]
        1 & \text{if}\ \Omega^h_P \in \Omega^h_\mathrm{bc}\ \text{and}\ y^+_\mathrm{ib} \leq y^+_\mathrm{lam} \\ [0.2cm]
        0 & \text{elsewhere}
    \end{array}
    \right.\,,
\end{equation}
where $y^+_\mathrm{ib}$ is the $y^+$ value evaluated with respect to the immersed boundary, and $y^+_\mathrm{lam} \approx 11.0$ is a constant commonly used as an approximate threshold between the viscous sublayer and the logarithmic region of the fluid boundary layer~\citep{bredberg2000}.

\paragraph{Treatment of production and dissipation of $k$}
Note that specific IB-induced source terms were added only to equations for transport of linear momentum~\eqref{eq:hfdibrans}$_1$ and of turbulence kinetic energy~\eqref{eq:hfdibkomega}$_1$. On the other hand, the immersed boundary also affects the values of production and dissipation of $k$, i.e., of $P_k$ and $\omega$, respectively. But, for $\omega$, this cannot be accounted for by introducing another IB-induced source term, since $\omega$ goes to infinity when approaching smooth walls~\citep{wilcox2006}. Instead, imposed fields $\omega_\mathrm{ib}$ and $P_\mathrm{ib}$ are calculated and assigned directly to $\omega$ and $P_k$ in cells with active mask field $\alpha_\omega$. This mask field is defined as
\begin{equation}
\label{eq:alphaomega}
    \alpha_\omega = \left\{
    \begin{array}{rl}
        1 & \text{if}\ \Omega^h_P \in (\Omega^h_\mathrm{ic} \cup \Omega^h_\mathrm{bc}) \\ [0.2cm]
        0 & \text{elsewhere}
    \end{array}
    \right.\,.
\end{equation}
For the production of $k$, the direct assignment of $P_\mathrm{ib}$ is straightforward, since no transport equation for $P_k$ is considered and $P_\mathrm{ib}$ is evaluated explicitly prior to the assembly of the turbulence model equations~\eqref{eq:hfdibkomega}. Conversely, the imposition of $\omega_\mathrm{ib}$ to the $\omega$ field is realized via a direct manipulation of the discretization matrix of the $\omega$ equation. In the affected cells, the $\omega$ equation is therefore not solved. Note that an analogous treatment, based on direct enforcement within the discretized system, is also employed in OpenFOAM to impose Dirichlet boundary conditions for $\omega$ near walls~\citep{oF}. Same treatment as for $\omega$ is employed for $\varepsilon$ in $\varepsilon$-based turbulence models.

\subsection{Calculation of $y^+$ at the immersed boundary}
\label{sub:yplus}
To switch between the variants of mask fields~\eqref{eq:alphauk}, it is necessary to compute $y^+_\mathrm{ib}$ at the immersed boundary, i.e., to decide if the boundary layer around IB is resolved. For a boundary cell $\Omega^h_P \in\Omega_{\mathrm{bc}}^h$, $y^+_\mathrm{ib}$ is computed as
\begin{equation}
\label{eq:yplus}
    y^+_\mathrm{ib} = \frac{y_\mathrm{eff}\ u_\tau}{\nu},\quad y_\mathrm{eff} = 0.5\,\left[y_\perp + 0.5\,\delta_\mathrm{cell}\right],
\end{equation}
where $y_\perp$ is the perpendicular distance from the cell center to the solid surface and $\delta_\mathrm{cell}$ is an estimate of the cell size. For meshes with hex cells, the cell size can be estimated as $\delta_\mathrm{cell} = V(\Omega_P^h)^{1/3}$, where $V(\Omega_P^h)$ is the cell volume. For highly distorted meshes, the size can be calculated from the cell bounding box constructed from the positions of its vertices and projecting the bounding box vector size $\bm{b}$ onto the surface normal direction yielding $\delta_\mathrm{cell} = |\bm{b}\cdot\bm{n}|$.

\begin{figure}[htbp]
    \centering
    \includegraphics{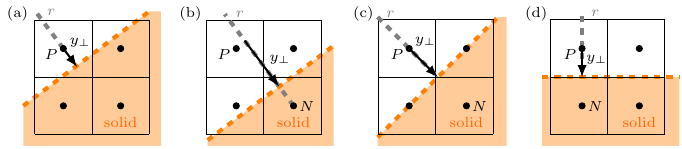}
    \caption{Visualization of $y_\perp$ for a boundary cell $\Omega^h_P$ in different situations with $\Omega^h_N$ as its in-solid neighbor. In all cases, orange indicates the solid surface position and $r$ the normal direction to the surface. (a) Situation where $\lambda_P \in (0,0.5)$. (b) Situation where $\lambda_P = 0$ and $\lambda_N < 1.0$. (c) Situation where $\lambda_P = 0$, $\lambda_N = 1.0$ and cells $\Omega^h_P$ and $\Omega^h_N$ share one vertex. (d) Same situation as in (c) but $\Omega^h_P$ and $\Omega^h_N$ share two vertices.}
    \label{fig:yperp}
\end{figure}
Now, the computation of $y^+_\mathrm{ib}$~\eqref{eq:yplus} depends on $y_\perp$, that is, on the perpendicular distance to the immersed boundary. For a boundary cell $\Omega^h_P\in\Omega_{\mathrm{bc}}^h$ and its in-solid neighbor $\Omega^h_N\in\Omega_{\mathrm{ic}}^h$ that shares at least one vertex with $\Omega^h_P$, $y_\perp$ is computed as
\begin{equation}
\label{eq:yortho}
    y_\perp = \left\{
    \begin{array}{cl}
        |\sigma_{\perp,P}| & \text{if}\ \lambda_P \in (0,0.5) \\ [0.3cm]
        \bm{n}_N \cdot \left[P - \left(N + |\sigma_{\perp,N}|\bm{n}_N\right)\right] & \text{if}\ \lambda_P = 0\ \text{and}\ \lambda_N \in \left[0.5, 1.0\right) \\ [0.3cm]
        \|P - \langle v_{PN} \rangle\| & \text{if}\ \lambda_P = 0\ \text{and}\ \lambda_N = 1.0
    \end{array}
    \right.,
\end{equation}
where $\langle v_{PN} \rangle$ is the average of the shared vertices and $\sigma$ and $\bm{n}$ are computed as
\begin{equation}
\label{eq:yorthoaux}
    \sigma_\perp = V(\Omega^h)^{1/3}\,\text{tanh}^{-1}\left(1 - 2\,\lambda \right),\quad \bm{n} = \frac{-\nabla \lambda}{\norm{\nabla \lambda}}\,.
\end{equation}
To find the in-solid neighbor $\Omega_N^h$, all neighbors of the boundary cell $\Omega_P^h$ are checked and the one with the lowest value of $-\bm{n}_P \cdot (P-N)$ is chosen, where $P$ and $N$ are the cell centers and $\bm{n}_P$ is the outer unit normal to the solid surface constructed in $P$. The corresponding $y_\perp$ computation procedure is illustrated in Figure~\ref{fig:yperp}.

\begin{figure}[htbp]
    \centering
    \includegraphics{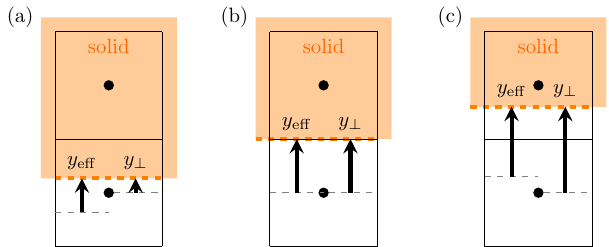}
    \caption{Comparison of $y_\mathrm{eff}$ and $y_\perp$ for three cases of solid surface position. In each case, the bottom cell is a boundary cell and the top an in-solid cell. (a) Boundary cell is intersected by the solid surface. (b) The solid surface goes in between the cell pair. (c) In-solid cell is intersected.}
    \label{fig:yeff}
\end{figure}
Nevertheless, the perpendicular distance $y_\perp$ is not used directly in the computation of $y^+_\mathrm{ib}$~\eqref{eq:yplus}. Instead, an effective perpendicular distance $y_\mathrm{eff}$ is utilized, as it more accurately represents the situation in which the boundary cell (or a neighboring in-solid cell) behaves as if intersected by the solid surface. This is illustrated in Figure~\ref{fig:yeff} for three representative cases of solid surface position.

Moreover, the use of $y_\mathrm{eff}$ leads to smoother variation of $y^+_\mathrm{ib}$. For instance, if cell pairs (a) and (c) from Figure~\ref{fig:yeff} are adjacent, the variation in $y_\mathrm{eff}$ is significantly smaller than that in $y_\perp$. Thus, employing $y_\mathrm{eff}$ translates into a smoother spatial distribution of $y^+_\mathrm{ib}$, an effect that becomes well visible for complex geometries. In~\ref{app:effandperp}\revs{, this is demonstrated in the simulation of flow over an Ahmed body, where the use of $y_\mathrm{eff}$ provided a better estimate of the recirculation region size and shape.}

\subsection{Calculation of imposed fields}
\label{sub:imval}
The imposed fields $\bm{u}_\mathrm{ib}$, $k_\mathrm{ib}$, $\omega_\mathrm{ib}$ or $\varepsilon_\mathrm{ib}$, and $P_\mathrm{ib}$ are calculated differently for the different cell groups, visualized in Figure~\ref{fig:lamomega}d. In the in-solid cells~$\Omega_{\mathrm{ic}}^h$, the imposed fields are prescribed as
\begin{equation}
\begin{array}{*1{>{\displaystyle}c}}
\label{eq:imval}
    \bm{u}_\mathrm{ib} = \bm{u}_\mathrm{s},\quad k_\mathrm{ib} = 0,\quad \omega_\mathrm{ib} = \displaystyle\max_{\Omega^h_\mathrm{bc}\, \cup\ \Omega^h_\mathrm{fc}}(\omega^\mathrm{old}),\\ \varepsilon_\mathrm{ib} = \displaystyle\max_{\Omega^h_\mathrm{bc}\, \cup\ \Omega^h_\mathrm{fc}}(\varepsilon^\mathrm{old}), \quad P_\mathrm{ib} = 0\,,
\end{array}
\end{equation}
where $\bm{u}_\mathrm{s}$ is the velocity of the solid and the superscript $"\mathrm{old}"$ denotes the previous iteration or the initial guess.

In the boundary cells~$\Omega_{\mathrm{bc}}^h$, the calculation of the imposed fields is based on the value of $y^+_\mathrm{ib}$. 
The values of $\bm{u}_\mathrm{ib}$ and $k_\mathrm{ib}$ are required only when $y^+_\mathrm{ib} \leq y^+_\mathrm{lam}$, since only then the corresponding mask fields are active in the boundary cells, see~\eqref{eq:alphauk}. The calculation itself is then done using polynomial interpolation.

\begin{figure}[htbp]
    \centering
    \includegraphics{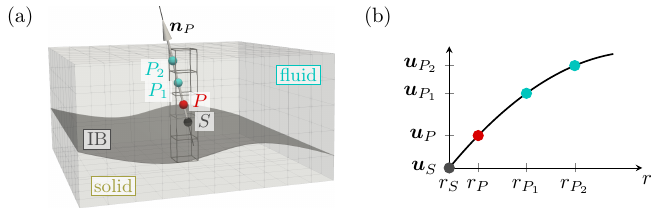}
    \caption{Construction of the interpolation polynomial for a boundary cell $\Omega^h_P$. (a) Interpolation points located along the direction of the solid surface normal $\bm{n}_P$. The points $P_1$ and $P_2$ are in the free stream, $S$ is on the solid surface, and $P$ in the boundary cell center. (b) Example of a quadratic interpolation profile. Adapted from~\citep{kubickova2023}.}
    \label{fig:interpolation}
\end{figure}
The interpolation polynomial is constructed in the same way as in openHFDIB~\citep{isoz2022, studenik2024}. The construction process is illustrated in Figure~\ref{fig:interpolation}. In particular, the interpolation is carried out in the normal direction to the solid surface, see Figure~\ref{fig:interpolation}a, and the polynomial, see the example in Figure~\ref{fig:interpolation}b, is constructed based on the values from the interpolation points located in the free stream and the value at the solid surface. More details on the interpolation process can be found in~\citep{isoz2022}.

For calculations in openHFDIBRANS, the values in the free stream are taken from the previous iteration or the initial guess. Regarding the surface values, the velocity is set equal to the velocity of the solid $\bm{u}_\mathrm{s}$ and $k$ is calculated using wall functions~\citep{kalitzin2005}. The utilized form of the wall functions is the one used in OpenFOAM~\citep{oF}, i.e.,
\begin{equation}
\label{eq:kws}
    k = k_\mathrm{vis}\,u_\tau^2,\quad k_\mathrm{vis} = \frac{2400\,C_f}{C^2_\mathrm{eps2}},\quad C_f = \frac{1}{(y^+_\mathrm{ib} + C)^2} + \frac{2\,y^+_\mathrm{ib}}{C^3} - \frac{1}{C^2}\,,
\end{equation}
where $C_\mathrm{eps2}$ and $C$ are model constants~\citep{wilcox2006} and $u_\tau$ is the friction velocity calculated as
\begin{equation}
\label{eq:frictionvel}
    u_\tau = C_{\nu}^{\,0.25}\, \sqrt{k_P}\,,
\end{equation}
where $C_\nu$ is another model constant, and $k_P$ is the value of $k$ in cell $\Omega^h_P$ from the previous iteration or the initial guess.

The values of the imposed fields $\omega_\mathrm{ib}$ or $\varepsilon_\mathrm{ib}$, and $P_\mathrm{ib}$ are needed regardless of the $y^+_\mathrm{ib}$ value. The corresponding mask field is always active, see~\eqref{eq:alphaomega}. The calculation of the imposed fields is done using wall functions~\citep{kalitzin2005} in the form used in OpenFOAM~\citep{oF}, which is
\begin{equation}
\label{eq:omegaws}
\begin{array}{rcl}
    \omega_\mathrm{ib} & = & \displaystyle\frac{u_\tau^2}{\nu}\cdot\left\{
    \begin{array}{rc}
        \displaystyle\frac{6}{\beta_1\,(y^+_\mathrm{ib})^2} & \text{if}\ \ y^+_\mathrm{ib} \leq y^+_\mathrm{lam}\\ [0.5cm]
        \displaystyle\frac{1}{\kappa\,\sqrt{C_\nu}\,y^+_\mathrm{ib}} & \text{if}\ \ y^+_\mathrm{ib} > y^+_\mathrm{lam}
    \end{array}
    \right.\!\!\!\!\quad,\\[1.0cm]
    \varepsilon_\mathrm{ib} & = & \displaystyle\frac{u_\tau^4}{\nu}\cdot\left\{
    \begin{array}{rc}
        \displaystyle\frac{2\,k}{u_\tau^2\,(y^+_\mathrm{ib})^2} & \text{if}\ \ y^+_\mathrm{ib} \leq y^+_\mathrm{lam}\\ [0.5cm]
        \displaystyle\frac{1}{\kappa\,y^+_\mathrm{ib}} & \text{if}\ \ y^+_\mathrm{ib} > y^+_\mathrm{lam}
    \end{array}
    \right.\!\!\!\!\quad,\\[1.0cm]
    P_\mathrm{ib} & = & \left\{
    \begin{array}{rc}
        P_k & \text{if}\ \ y^+_\mathrm{ib} \leq y^+_\mathrm{lam}\\ [0.5cm]
        \displaystyle\frac{\kappa\,\nu\,y^+_\mathrm{ib}\,\norm{\bm{n}_\mathrm{ib}\cdot\nabla\bm{u}}^2}{\log^2(E\,y^+_\mathrm{ib})} & \text{if}\ \ y^+_\mathrm{ib} > y^+_\mathrm{lam}
    \end{array}
    \right.\!\!\!\!\quad,
\end{array}
\end{equation}
with $\beta_1$, $\kappa$ and $E$ being model constants~\citep{wilcox2006}.

\begin{figure}[htbp]
    \centering
    \includegraphics{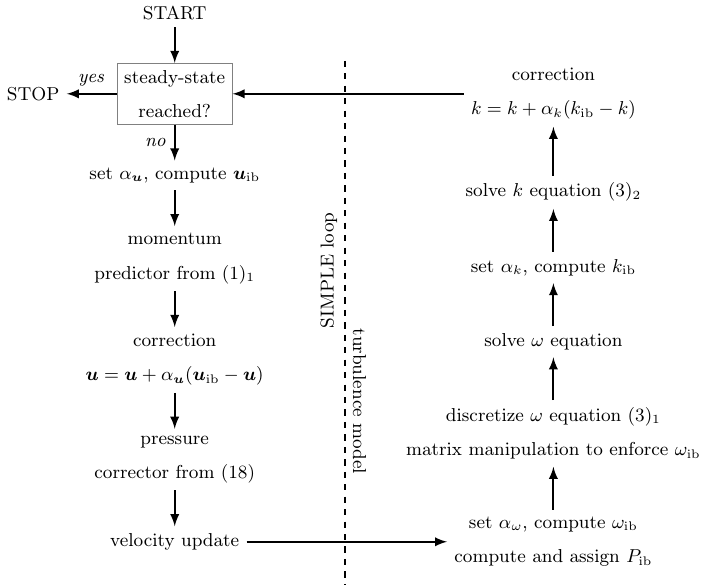}
    \caption{Visualization of the SIMPLE algorithm with additional steps required in openHFDIBRANS.}
    \label{fig:simplehfdibrans}
\end{figure}

\subsubsection{SIMPLE-based solution strategy with immersed boundary enforcement}
\label{par:simplehfdibrans}
In openHFDIBRANS, the RANS equations~\eqref{eq:hfdibrans} and the turbulence model equations~\eqref{eq:hfdibkomega} are solved using the SIMPLE algorithm~\citep{patankar1972} that is extended by additional steps. The resulting solver loop, illustrated for the $k$-$\omega$ model in Figure~\ref{fig:simplehfdibrans}, is described below.

First, before the momentum predictor~\citep{moukalled2016}, the values of $\bm{u}_\mathrm{ib}$ are calculated. Then the momentum equation is solved and a correction on the resulting velocity field is applied as
\begin{equation}
\label{eq:corru}
    \bm{u} = \bm{u} + \alpha_{\bm{u}}(\bm{u}_\mathrm{ib} - \bm{u})\,.
\end{equation}
With this correction, the values of $\bm{u}_\mathrm{ib}$ are directly imposed in cells where the mask field $\alpha_{\bm{u}}$ is active. This correction is important for achieving convergence in steady-state, as the system matrix must be under-relaxed to ensure diagonal dominance~\citep{oF}. Under-relaxation adds auxiliary source terms evaluated using the velocity field from the previous iteration. The correction~\eqref{eq:corru} ensures that these source terms remain consistent with the imposed values. Then the pressure corrector equation is constructed from the modified continuity equation~\eqref{eq:hfdibrans}$_2$. Let $M$ be the linear matrix of the momentum equation and $M = D + L + U$ its decomposition to diagonal $D$, $L$ lower triangular and $U$ upper triangular matrices. Considering these, the pressure corrector equation can be written as
\begin{equation}
    \label{eq:prescorr}
    \nabla\cdot(D^{-1}\,\nabla\tilde{p}) = \nabla\cdot(D^{-1}\bm{f}_\mathrm{ib}) - \nabla\cdot\left[D^{-1}(L + U)\bm{u}^*\right] - \nabla\cdot\left(\lambda\bm{u}^*\right)\,,
\end{equation}
where $\bm{u}^*$ denotes the velocity acquired from the momentum-equation solution. The equation retains the same form throughout the computational domain, including both fluid and solid regions. However, the last term effectively restricts the incompressibility constraint to the fluid region, thus avoiding conflict with the immersed-boundary forcing term $\bm{f}_\mathrm{ib}$. After the pressure corrector step, the velocity is updated as in the standard SIMPLE algorithm~\citep{moukalled2016}.

After updating the velocity, the solver proceeds to the turbulence model equations. First, the imposed fields $\omega_\mathrm{ib}$ and $P_\mathrm{ib}$ are evaluated, and $P_\mathrm{ib}$ is assigned to $P_k$. The $\omega$ equation is then discretized, with $\omega_\mathrm{ib}$ enforced via direct modification of the discretization matrix; the equation is subsequently solved and $k_\mathrm{ib}$ is computed. Finally, as in the velocity treatment, a correction of the $k$ field is applied, thereby completing the solver iteration.

To summarize, the additional operations required by openHFDIBRANS include
\begin{inparaenum}[(i)]
    \item{setting of mask fields $\alpha_{\bm{u}}$, $\alpha_\omega$ and $\alpha_k$,}
    \item{computation of imposed fields $\bm{u}_\mathrm{ib}$, $k_\mathrm{ib}$, $\omega_\mathrm{ib}$ and $P_\mathrm{ib}$,}
    \item{application of corrections on solved $\bm{u}$ and $k$ fields, and}
    \item{manipulation with the discretization matrix of the $\omega$ equation.}
\end{inparaenum}

\section{Results and Discussion}
\label{sec:res}
The openHFDIBRANS solver was developed to be an effective tool for geometry optimization combining immersed boundary method, steady-state solution algorithm and RANS turbulence models. The solver is open-source and available with simulation setups at \href{https://github.com/techMathGroup/openHFDIBRANS}{https://github.com/techMathGroup/openHFDIBRANS}.

Six test cases of increasing geometric complexity were used to assess the solver capabilities:
\begin{inparaenum}[(i)]
    \item{two-dimensional (2D) pipe flow,}
    \item{3D pipe flow,}
    \item{2D flow over a backward-facing step,}
    \item{2D flow over a smooth cylinder,}
    \item{2D flow over a NACA-0009 profile, and}
    \item{3D flow over an Ahmed body.}
\end{inparaenum}
At the section end, the usage of openHFDIBRANS in an engineering application is presented featuring a simulation of coolant flow in a die-casting die core. In all presented cases, openHFDIBRANS was compared with the standard geometry-conforming solver simpleFoam available from OpenFOAM~\citep{oF}, and where available, also with experimental data and data from other immersed boundary solvers, \revs{see Table~\ref{tab:availsum}.}

\begin{table}[htbp]
    \centering
    \small
    \revs{
    \begin{tabular}{c|ccccccc}
    data source & \rotatebox{90}{2D pipe flow} & \rotatebox{90}{3D pipe flow} & \rotatebox{90}{backward-facing step} & \rotatebox{90}{cylinder} & \rotatebox{90}{NACA-0009} & \rotatebox{90}{Ahmed body} & \rotatebox{90}{die-casting die core} \\
    \toprule
    openHFDIBRANS        & \cmark & \cmark & \cmark & \cmark & \cmark & \cmark & \cmark \\
    simpleFoam           & \cmark & \cmark & \cmark & \cmark & \cmark & \cmark & \cmark \\
    experiment/reference & \xmark & \xmark & \cmark & \cmark & \cmark & \cmark & \xmark \\
    other IB solver & \xmark & \xmark & \cmark & \cmark & \cmark & \cmark & \xmark \\
    \end{tabular}
    }
    \caption{\revs{Availability of data sources for each studied case.}}
    \label{tab:availsum}
\end{table}

Collectively, the testing examined different aspects of openHFDIBRANS including:
\begin{inparaenum}[(i)]
    \item{sensitivity to Reynolds number, turbulence model selection, and interface position (2D pipe flow),}
    \item{sensitivity to mesh and interface distortion (3D pipe flow),}
    \item{calculation of wall shear stress, skin friction coefficients and reattachment length (backward-facing step),}
    \item{evaluation of drag and lift coefficients (smooth cylinder, NACA-0009 profile), and}
    \item{characterization of a recirculation region behind a bluff body (Ahmed body).}
\end{inparaenum}

Moreover, additional studies were added to the Appendix focusing on:
\begin{inparaenum}[(i)]
    \item{effect of different approaches to $y^+_\mathrm{ib}$ evaluation (\ref{app:effandperp}),}
    \item{sensitivity to selected numerical schemes (\ref{app:numschemes}),}
    \item{solver behavior while changing misalignment of mesh faces and immersed boundary (\ref{app:misalign}),}
    \item{mesh-refinement study (\ref{app:meshref}),}
    \item{study of convergence, matrix condition numbers and solenoidality of velocity field (\ref{app:stiffandconv}), and}
    \item{scaling in parallel computations (\ref{app:parallel}).}
\end{inparaenum}

\begin{figure}[htbp]
    \centering
    \includegraphics{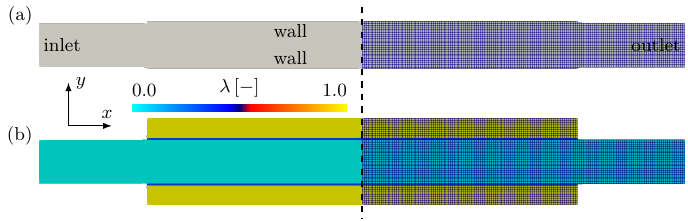}
    \caption{Computational domains and meshes used for simulations of 2D pipe flow. Dashed black line indicates the sampling line used for result comparison. (a) Domain and mesh used for simpleFoam with indicated boundaries. (b) Domain and mesh used for openHFDIBRANS colored by $\lambda$ field.}
    \label{fig:smeshes}
\end{figure}

\subsection{Sensitivity to Reynolds number, turbulence model and interface position in two-dimensional pipe flow}
\label{sec:straight}
In the test case with the simplest geometry, 2D pipe flow, openHFDIBRANS maintained consistent accuracy for a wide range of flow Reynolds numbers, different turbulence models, and variations in the solid-fluid interface position.

\paragraph{Simulation setup}
\label{par:simsetpipe}
For both the openHFDIBRANS and simpleFoam solvers, equivalent computational domains and meshes were used, see Figure~\ref{fig:smeshes}, with the pipe wall represented by an immersed boundary for openHFDIBRANS and a geometry-conforming mesh for simpleFoam. Note that the computational domains were formally three-dimensional, but comprised a single cell in the out-of-plane direction.

At the pipe inlet, the flow was prescribed with a zero-gradient boundary condition for pressure, and a Dirichlet boundary condition for velocity $\bm{u} = (u_\mathrm{in},0,0)^\mathrm{T}$ and turbulence variables. The inlet values of the turbulence variables were calculated according to~\citep{russo2016} as
\begin{equation}
    \label{eq:bcspkoein}
    k_\mathrm{in} = \frac{3}{2}\,u_\mathrm{in}^2\,I_\mathrm{in}^2,\quad \varepsilon_\mathrm{in} = C_\nu\,\frac{k_\mathrm{in}^\frac{3}{2}}{\ell_\mathrm{t}},\quad \omega_\mathrm{in} = \frac{\varepsilon_\mathrm{in}}{C_\nu\,k_\mathrm{in}}\,,
\end{equation}
where $I_\mathrm{in} = 2.0\,\%$ was the turbulence intensity and $\ell_\mathrm{t}$ was the turbulent length-scale determined as a tenth of the pipe inlet width.

At the pipe outlet, a Dirichlet boundary condition for pressure $\tilde{p}_\mathrm{out} = 0$, and a zero-gradient boundary condition for velocity and turbulence variables were applied. For the walls, the zero-gradient condition for pressure, the no-slip boundary condition for velocity, and wall functions for turbulence variables were used. In openHFDIBRANS, walls within the solid phase, identified by $\lambda > 0$ in adjacent cells, were treated by imposing zero-gradient boundary conditions for all variables.

Both simpleFoam and openHFDIBRANS were run using the same numerical settings. These settings were chosen to enhance the stability of the simulations, including the upwind scheme for discretizating the divergence term in all equations and strong under-relaxation (by a factor of 0.85). Consequently, the same numerical settings could be retained across all test cases, ensuring consistent numerical behavior and enabling direct comparison of the results.

A study of \revs{the} effect of different numerical schemes for divergence terms is provided in~\ref{app:numschemes} for the backward-facing step benchmark, flow over a smooth cylinder and flow over a NACA-0009 airfoil. When compared with simpleFoam, openHFDIBRANS did not exhibit over-sensitivity to \revs{the} chosen numerical scheme. \revs{Both simpleFoam and openHFDIBRANS achieved higher accuracy with high-order schemes, but also showed instabilities in some cases, e.g., for the \textit{vanLeer} scheme in the simulation of flow over a NACA-0009 airfoil.}

\begin{figure}[htbp]
    \centering
    \includegraphics{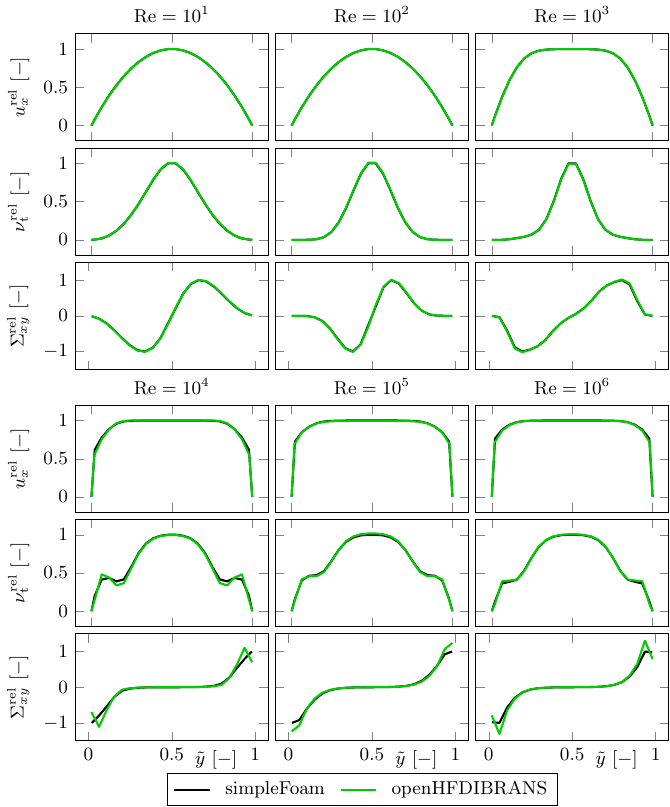}
    \caption{Simulation of 2D pipe flow with different $\mathrm{Re}$ and the $k$-$\omega$ turbulence model. Compared are the $x$ velocity component ($u_x$), turbulent viscosity ($\nu_\mathrm{t}$), and the $x$-$y$ component of the Reynolds stress tensor divided by the fluid density ($\Sigma_{xy}$). Data were sampled along the black line depicted in Figure~\ref{fig:smeshes}. All variables were normalized with respect to simpleFoam (sF) as $\varphi^\mathrm{rel} = \left[\varphi - \min{(\varphi_\mathrm{sF})}\right]/\left[\max{\varphi_\mathrm{sF}
    )} - \min{(\varphi_\mathrm{sF})}\right]$.}
    \label{fig:straight}
\end{figure}
\subsubsection{Sensitivity to Reynolds number}
\label{sub:reynum}
For Reynolds numbers ranging from $10^1$ to $10^6$, the openHFDIBRANS results closely matched those of simpleFoam, see Figure~\ref{fig:straight}. The flow Reynolds number was varied by adjusting the inlet velocity $u_\mathrm{in}$, while the fluid kinematic viscosity remained fixed at $\nu = 10^{-6}\,\mathrm{m}^{2}\,\mathrm{s}^{-1}$. For all tested $\mathrm{Re}$, the same computational meshes were used. Consequently, the $y^+$ value ranged from $y^+ \ll 1$ for low $\mathrm{Re}$ to $y^+ \lesssim 10^3$ for high $\mathrm{Re}$.

\begin{table}[htbp]
    \centering
    \small
    \begin{tabular}{c|cccccc}
    $\mathrm{Re}$ & $10^1$ & $10^2$ & $10^3$ & $10^4$ & $10^5$ & $10^6$ \\
    \toprule
    $\text{avg}\!\left[\ell^2(u^\mathrm{rel}_x)\right]$ & 9.630e-04 & 9.367e-04 & 9.481e-04 & 1.997e-02 & 8.821e-03 & 1.426e-02 \\
    $\sum \ell^2(u^\mathrm{rel}_x)$ & 4.517e-03 & 4.394e-03 & 4.447e-03 & 9.366e-02 & 4.138e-02 & 6.688e-02 \\
    $\text{avg}\!\left[\ell^2(\nu^\mathrm{rel}_\mathrm{t})\right]$ & 2.976e-03 & 4.076e-03 & 4.054e-03 & 3.525e-02 & 1.185e-02 & 1.364e-02 \\
    $\sum \ell^2(\nu^\mathrm{rel}_\mathrm{t})$ & 1.396e-02 & 1.912e-02 & 1.902e-02 & 1.653e-01 & 5.558e-02 & 6.396e-02 \\
    $\text{avg}\!\left[\ell^2(\Sigma^\mathrm{rel}_\mathrm{xy})\right]$ & 9.075e-03 & 1.541e-02 & 1.510e-02 & 1.375e-01 & 8.469e-02 & 1.123e-01 \\
    $\sum \ell^2(\Sigma^\mathrm{rel}_\mathrm{xy})$ & 4.256e-02 & 7.226e-02 & 7.085e-02 & 6.451e-01 & 3.973e-01 & 5.269e-01 \\
    \end{tabular}
    \caption{Average and total $\ell^2$ norm calculated from the difference of profiles given in Figure~\ref{fig:straight}.}
    \label{tab:straightnorm}
\end{table}
To quantify the agreement, the $\ell^2$ norm of the difference between the profiles was computed, showing mostly consistent error values, see Table~\ref{tab:straightnorm}. The largest error for all monitored fields was observed at $\mathrm{Re} = 10^4$. In this case, the $y^+$ increased along the pipe, with a minimum at $\sim\!2$ and a maximum at $\sim\!16$, traversing a buffer region between the viscous sublayer and the logarithmic region. This is reflected by a switch in the wall function formulations used in both simpleFoam~\citep{kalitzin2005} and openHFDIBRANS~\eqref{eq:omegaws}. However, in openHFDIBRANS, also the mask fields for $\bm{u}$ and $k$ change from active to inactive~\eqref{eq:alphauk}, likely causing the higher error value. Despite the non-ideal treatment of the buffer region, the error magnitudes remained small throughout the entire range of $\mathrm{Re}$. In addition, a possible remedy is the introduction of a blending function to smooth the transition of the mask fields.

\begin{figure}[htbp]
    \centering
    \includegraphics{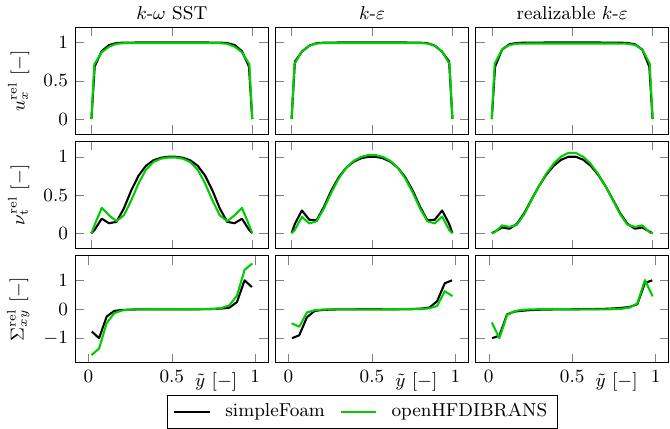}
    \caption{Simulation of 2D pipe flow with $\mathrm{Re} = 10^6$ and different turbulence models. Data sampling and normalization was done as in Figure~\ref{fig:straight}.}
    \label{fig:moretm}
\end{figure}

\begin{table}[htbp]
    \centering
    \small
    \begin{tabular}{c|ccc}
    turbulence model & $k$-$\omega$ SST & $k$-$\varepsilon$ & realizable $k$-$\varepsilon$ \\
    \toprule
    $\text{avg}\!\left[\ell^2(u^\mathrm{rel}_x)\right]$ & 1.627e-02 & 1.958e-02 & 2.623e-02 \\
    $\sum \ell^2(u^\mathrm{rel}_x)$ & 7.633e-02 & 9.185e-02 & 1.230e-01 \\
    $\text{avg}\!\left[\ell^2(\nu^\mathrm{rel}_\mathrm{t})\right]$ & 8.158e-02 & 4.007e-02 & 2.496e-02 \\
    $\sum \ell^2(\nu^\mathrm{rel}_\mathrm{t})$ & 3.826e-01 & 1.879e-01 & 1.171e-01 \\
    $\text{avg}\!\left[\ell^2(\Sigma^\mathrm{rel}_\mathrm{xy})\right]$ & 2.812e-01 & 1.903e-01 & 1.706e-01 \\
    $\sum \ell^2(\Sigma^\mathrm{rel}_\mathrm{xy})$ & 1.319e+00 & 8.924e-01 & 8.000e-01 \\
    \end{tabular}
    \caption{Average and total $\ell^2$ norm calculated from the difference of profiles given in Figure~\ref{fig:moretm}.}
    \label{tab:moretmnorm}
\end{table}

\begin{figure}[htbp]
    \centering
    \includegraphics{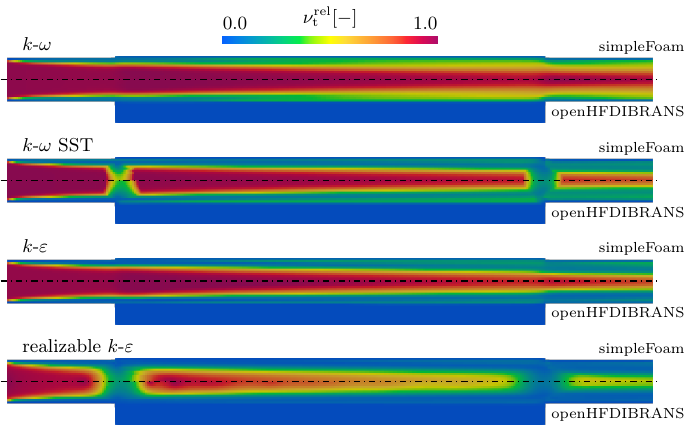}
    \caption{Qualitative comparison of $\nu_\mathrm{t}$ fields from the 2D simulation of pipe flow run with different turbulence models.}
    \label{fig:moretmqua}
\end{figure}
\subsubsection{Performance with different turbulence models}
\label{sub:otm}
With other turbulence models ($k$-$\omega$ SST, $k$-$\varepsilon$ and realizable $k$-$\varepsilon$), openHFDIBRANS remained consistent with simpleFoam in the predicted velocity, but showed slightly larger errors in $\nu_\mathrm{t}$ and $\Sigma_{xy}$, see Figure~\ref{fig:moretm} and Table~\ref{tab:moretmnorm}. A qualitative comparison of the $\nu_\mathrm{t}$ fields revealed sharp changes for the $k$-$\omega$ SST and realizable $k$-$\varepsilon$ models. In the openHFDIBRANS solution, these transitions were slightly shifted downstream, see Figure~\ref{fig:moretmqua}. Nevertheless, the overall distribution of $\nu_\mathrm{t}$ showed the same trends as the simpleFoam solution, and the observed discrepancies had minimal impact on velocity.

\begin{figure}[htbp]
    \centering
    \includegraphics{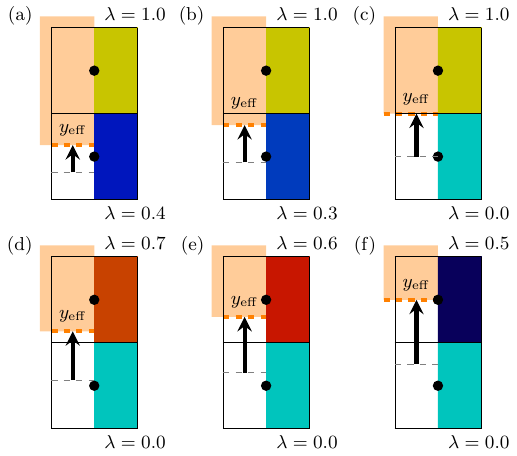}
    \caption{Investigated variations of solid-fluid interface position. On the left side, the position of the solid surface (orange) and the effective distance $y_\mathrm{eff}$ are depicted. On the right side, the resulting $\lambda$ field values are given. The situation (b) matches the state of interface position depicted in Figure~\ref{fig:smeshes}.}
    \label{fig:laminterface}
\end{figure}

\begin{figure}[htbp]
    \centering
    \includegraphics{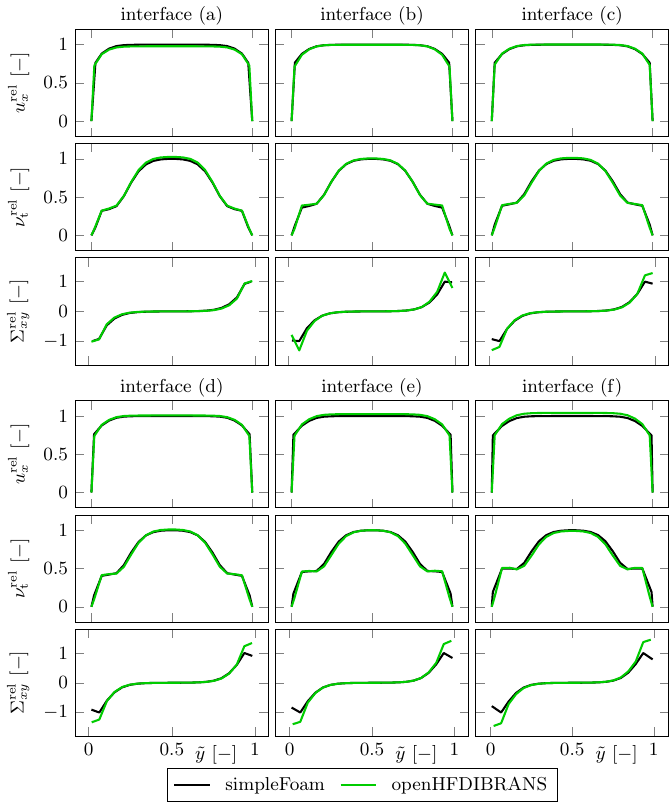}
    \caption{Simulation of 2D pipe flow with different variations in the solid-fluid interface positions, $\mathrm{Re} = 10^6$ and $k$-$\omega$ model. The interface designations match those from Figure~\ref{fig:laminterface}. Data sampling and normalization was done as in Figure~\ref{fig:straight}.}
    \label{fig:difflam}
\end{figure}

\begin{table}[htbp]
    \centering
    \small
    \begin{tabular}{c|cccccc}
    interface & (a) & (b) & (c) & (d) & (e) & (f) \\
    \toprule
    $\text{avg}\!\left[\ell^2(u^\mathrm{rel}_x)\right]$ & 1.861e-02 & 1.426e-02 & 8.570e-03 & 1.104e-02 & 2.392e-02 & 3.709e-02 \\
    $\sum \ell^2(u^\mathrm{rel}_x)$ & 8.730e-02 & 6.688e-02 & 4.019e-02 & 5.179e-02 & 1.122e-01 & 1.740e-01 \\
    $\text{avg}\!\left[\ell^2(\nu^\mathrm{rel}_\mathrm{t})\right]$ & 1.706e-02 & 1.364e-02 & 1.590e-02 & 1.949e-02 & 3.266e-02 & 4.509e-02 \\
    $\sum \ell^2(\nu^\mathrm{rel}_\mathrm{t})$ & 8.001e-02 & 6.396e-02 & 7.459e-02 & 9.143e-02 & 1.532e-01 & 2.115e-01 \\
    $\text{avg}\!\left[\ell^2(\Sigma^\mathrm{rel}_\mathrm{xy})\right]$ & 1.771e-02 & 1.123e-01 & 1.262e-01 & 1.463e-01 & 1.898e-01 & 2.193e-01 \\
    $\sum \ell^2(\Sigma^\mathrm{rel}_\mathrm{xy})$ & 8.305e-02 & 5.269e-01 & 5.919e-01 & 6.862e-01 & 8.902e-01 & 1.029e+00 \\
    \end{tabular}
    \caption{Average and total $\ell^2$ norm calculated from the difference of profiles given in Figure~\ref{fig:difflam}.}
    \label{tab:difflamnorm}
\end{table}
\subsubsection{Sensitivity to position of solid-fluid interface}
\label{sub:difflam}
When changing the position of the solid-fluid interface, openHFDIBRANS achieved the highest accuracy when the immersed boundary was well aligned with mesh faces, while keeping acceptable error levels for less favorable interface positions, see Figure~\ref{fig:difflam}. The different interface variations that were considered are illustrated in Figure~\ref{fig:laminterface}. For openHFDIBRANS, the interface variations resulted in different values of the $\lambda$ field, but the original domain from Figure~\ref{fig:smeshes}b was still used. For simpleFoam, the width of the middle part of the original domain from Figure~\ref{fig:smeshes}a was adjusted to match the variations in interface position.

The better aligned interfaces (b)-(d) produced lower $\ell_2$ norms for $u_x$ and $\Sigma_{xy}$ and mostly also for $\nu_\mathrm{t}$. For worse aligned interfaces (a), (e) and (f), the openHFDIBRANS estimates were burdened with a higher error. Still, the overall robustness of openHFDIBRANS with respect to the position of IB; and, consequently, to variations in $y_\mathrm{eff}$ was satisfactory.

\begin{figure}[htbp]
    \centering
    \includegraphics{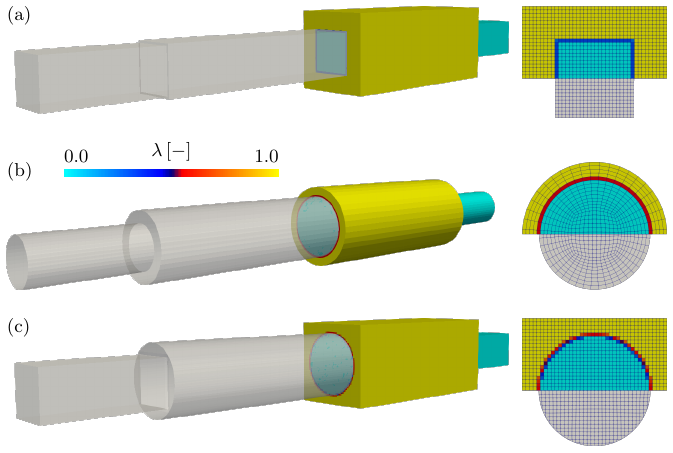}
    \caption{Isometric (left) and cross-sectional (right) view onto computational domains and meshes used for simulation of 3D pipe flow. Only halves of each domain are depicted. Domains used for simpleFoam are colored in grey and for openHFDIBRANS by $\lambda$ field.}
    \label{fig:meshes3D}
\end{figure}

\begin{figure}[htbp]
    \centering
    \includegraphics{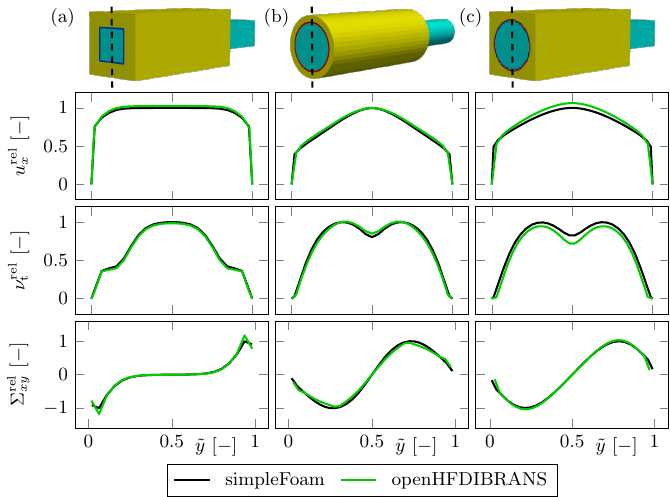}
    \caption{Simulation of 3D pipe flow with domains from Figure~\ref{fig:meshes3D}, $\mathrm{Re} = 10^6$ and $k$-$\omega$ turbulence model. For each domain, the data sampling line is given above the graphs. Data normalization was done as in Figure~\ref{fig:straight}.}
    \label{fig:pipes3D}
\end{figure}

\subsection{Effect of mesh and interface distortion in three-dimensional pipe flow}
\label{sub:pipe3D}
Three-dimensional pipe flow contained more severe mesh and interface distortion than the 2D case, but openHFDIBRANS maintained a comparable level of accuracy.

\paragraph{Simulation setup}
\label{par:simsetpipe3D}
From the design of the 2D pipe in Figure~\ref{fig:smeshes}, three 3D variants were created, see Figure~\ref{fig:meshes3D}. The first variant was generated by axial extrusion, the second by rotational extrusion, and the third by projecting the circular pipe of the second variant onto the first. Aside from the change in domains, the numerical settings remained identical to those of the 2D pipe flow.

\begin{table}[htbp]
    \centering
    \small
    \begin{tabular}{c|ccc}
    3D pipe variation & (a) & (b) & (c) \\
    \toprule
    $\text{avg}\!\left[\ell^2(u^\mathrm{rel}_x)\right]$ & 2.279e-02 & 2.115e-02 & 4.866e-02 \\
    $\sum \ell^2(u^\mathrm{rel}_x)$ & 1.069e-01 & 1.139e-01 & 2.665e-01 \\
    $\text{avg}\!\left[\ell^2(\nu^\mathrm{rel}_\mathrm{t})\right]$ & 1.700e-02 & 3.216e-02 & 7.489e-02 \\
    $\sum \ell^2(\nu^\mathrm{rel}_\mathrm{t})$ & 7.972e-02 & 1.732e-01 & 4.102e-01 \\
    $\text{avg}\!\left[\ell^2(\Sigma^\mathrm{rel}_\mathrm{xy})\right]$ & 6.560e-02 & 6.662e-02 & 5.934e-02 \\
    $\sum \ell^2(\Sigma^\mathrm{rel}_\mathrm{xy})$ & 3.077e-01 & 3.587e-01 & 3.250e-01 \\
    \end{tabular}
    \caption{Average and total $\ell^2$ norm calculated from the difference of profiles given in Figure~\ref{fig:pipes3D}.}
    \label{tab:pipes3Dnorm}
\end{table}

\paragraph{Effect of mesh and interface distortion}
\label{par:res3D}
For all the tested 3D variants, the agreement between simpleFoam and openHFDIBRANS was as strong as it was for the 2D case, see Figure~\ref{fig:pipes3D} and Table~\ref{tab:pipes3Dnorm}. The largest deviations occurred in the third variant, which had the worst alignment between the immersed boundary and mesh faces, see Figure~\ref{fig:meshes3D}c (right). Even with the worst-alignment, the $\ell^2$ norms of the errors remained within the same order of magnitude, showing that openHFDIBRANS scales well from simple 2D cases to more complex 3D geometries.

A more detailed study of openHFDIBRANS behavior with respect to the misalignment of immersed boundary and mesh faces is given in~\ref{app:misalign}. \revs{For misalignment angles ranging from $\beta = 0^\circ$ (no misalignment) to $\beta = 45^\circ$ (maximum misalignment), openHFDIBRANS showed no substantial variations in the solution.}

\begin{figure}[htbp]
    \centering
    \includegraphics{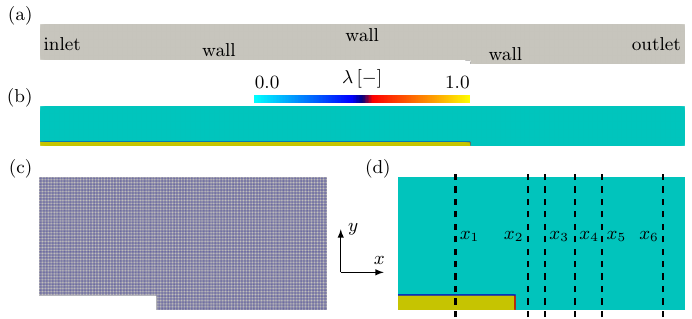}
    \caption{Backward-facing step benchmark. (a) Computational domain used by simpleFoam with indicated boundaries. (b) Computational domain and $\lambda$ field used by openHFDIBRANS. (c) Mesh in the domain from (a) in the vicinity of the step. (d) $\lambda$ field in the vicinity of the step with highlighted data sampling lines.}
    \label{fig:bfsmesh}
\end{figure}

\begin{figure}[htbp]
    \centering
    \includegraphics{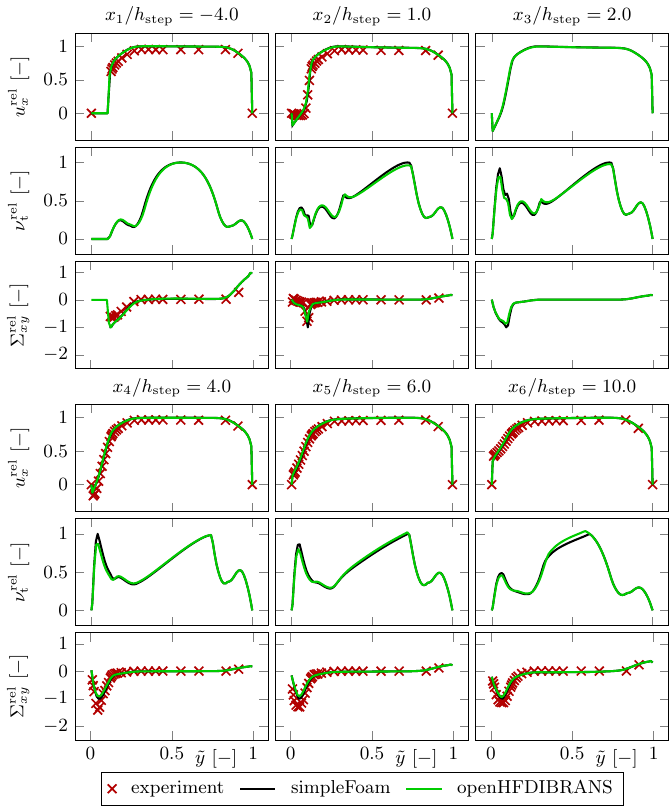}
    \caption{Backward-facing step benchmark. Profiles were sampled along lines given in Figure~\ref{fig:bfsmesh}d. Data were normalized as in Figure~\ref{fig:straight}. The step position is at $x/h_\mathrm{ste} = 0.0$.}
    \label{fig:bfs}
\end{figure}

\subsection{Flow over a backward-facing step}
\label{sec:bfs}
The backward-facing step benchmark features sudden boundary layer separation and reattachment, with openHFDIBRANS reproducing both phenomena with accuracy comparable to simpleFoam. Moreover, when the boundary layer was sufficiently resolved, both solvers closely matched the experimental data of~\citet{driver1985}.

\paragraph{Simulation setup}
\label{par:simsetbfs}
A detailed description of the benchmark can be found at NASA Turbulence Modeling Resource (TMR)~\citep{nasa}. The computational domains of simpleFoam and openHFDIBRANS were made as similar as possible, with the step represented by the immersed boundary in openHFDIBRANS, see Figure~\ref{fig:bfsmesh}.

The boundary conditions were prescribed identically to the 2D pipe flow case, see Section~\ref{sec:straight}, and the simulation conducted with the $k$-$\omega$ SST turbulence model. In accordance with NASA TMR~\citep{nasa}, the fluid kinematic viscosity was $\nu = 1.56\cdot10^{-5}\ \mathrm{m}^2\,\mathrm{s}^{-1}$, with the inlet velocity $u_\mathrm{in}$ set to ensure $\mathrm{Re} = 36\,000$ based on the step height of $h_\mathrm{step} = 0.0127\ \mathrm{m}$.

From the simulation results, the wall shear stress $\tau_\mathrm{w}$ and skin friction coefficient $C_\mathrm{f}$ were calculated as
\begin{equation}
    \tau_\mathrm{w} = \left(\nu+\nu_\mathrm{t}\right) \frac{\partial u_x}{\partial y}\big|_{y = y_\mathrm{wall}},\quad C_\mathrm{f} = 2\,\tau_\mathrm{w}/u_\mathrm{in}^2\,,
\end{equation}
where \textit{wall} stands for the bottom wall that comprises the backward-facing step. In openHFDIBRANS, parts of the \textit{wall} are represented by the immersed boundary.

\begin{figure}[htbp]
    \centering
    \includegraphics{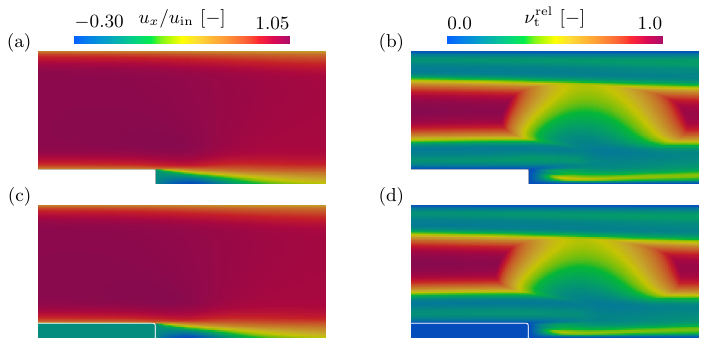}
    \caption{Zoomed view onto flow fields in the backward-facing step benchmark. (a-b) Fields from simpleFoam. (c-d) Fields from openHFDIBRANS where white indicates the position of the solid surface.}
    \label{fig:bfsqua}
\end{figure}

\begin{table}[htbp]
    \centering
    \small
    \begin{tabular}{c|cccccc}
    $x/h_\mathrm{step}$ & $-4.0$ & $1.0$ & $2.0$ & $4.0$ & $6.0$ & $10.0$ \\
    \toprule
    $\text{avg}\!\left[\ell^2(u^\mathrm{rel}_x)\right]$ & 1.035e-02 & 6.931e-03 & 7.362e-03 & 4.800e-03 & 5.928e-03 & 3.210e-03 \\
    $\sum \ell^2(u^\mathrm{rel}_x)$ & 9.139e-02 & 6.465e-02 & 6.867e-02 & 4.478e-02 & 5.529e-02 & 2.994e-02 \\
    $\text{avg}\!\left[\ell^2(\nu^\mathrm{rel}_\mathrm{t})\right]$ & 1.384e-02 & 2.264e-02 & 2.759e-02 & 2.637e-02 & 2.311e-02 & 3.062e-02 \\
    $\sum \ell^2(\nu^\mathrm{rel}_\mathrm{t})$ & 1.222e-01 & 2.112e-01 & 2.573e-01 & 2.459e-01 & 2.155e-01 & 2.856e-01 \\
    $\text{avg}\!\left[\ell^2(\Sigma^\mathrm{rel}_\mathrm{xy})\right]$ & 2.118e-02 & 2.629e-02 & 2.862e-02 & 2.361e-02 & 2.308e-02 & 1.682e-02 \\
    $\sum \ell^2(\Sigma^\mathrm{rel}_\mathrm{xy})$ & 1.871e-01 & 2.452e-01 & 2.670e-01 & 2.202e-01 & 2.153e-01 & 1.569e-01 \\
    \end{tabular}
    \caption{Average and total $\ell^2$ norm calculated from the difference of profiles given in Figure~\ref{fig:bfs}.}
    \label{tab:bfsnorm}
\end{table}

\paragraph{\revs{Comparison with simpleFoam}}
\label{par:resbfsexp}
At all sampling locations, openHFDIBRANS results \revs{were} in close agreement with simpleFoam in both $u_x$ and $\Sigma_{xy}$, see Figure~\ref{fig:bfs} and Table~\ref{tab:bfsnorm}. For $\nu_\mathrm{t}$, openHFDIBRANS results \revs{contained} small differences downstream of the step. A close-up view of the $u_x$ and $\nu_\mathrm{t}$ fields, see Figure~\ref{fig:bfsqua}, showed that there were sharp changes in $\nu_\mathrm{t}$ that were a little bit shifted in the openHFDIBRANS results. Nonetheless, the impact on the velocity field remained minimal.

\begin{figure}[htbp]
    \centering
    \includegraphics{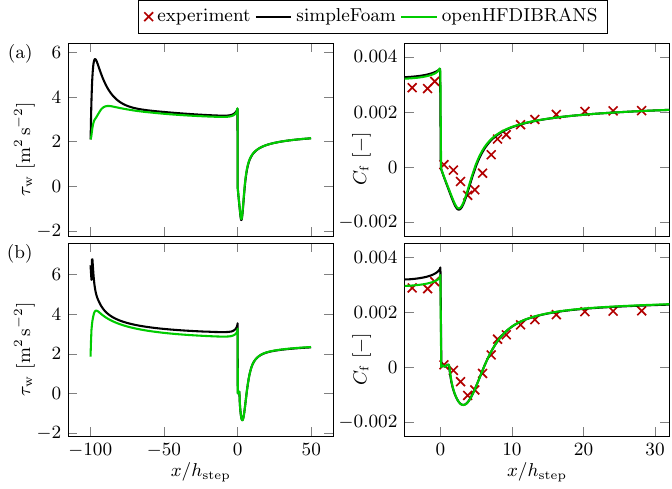}
    \caption{Evolution of the wall shear stress $\tau_\mathrm{w}$ and skin friction coefficient $C_\mathrm{f}$ sampled along the bottom wall in the backward-facing step benchmark simulated with mesh having (a) $\max(y_\mathrm{step}^+) = 70$ and (b) $\max(y_\mathrm{step}^+) = 15$.}
    \label{fig:bfswss}
\end{figure}

\begin{table}[htbp]
    \centering
    \small
    \begin{tabular}{c|cccc}
    mesh: position & (a): $x$ < 0.0 & (a): x > 0.0 & (b): x < 0.0 & (b): x > 0.0 \\
    \toprule
    $\text{avg}\!\left[\ell^2(u^\mathrm{rel}_x)\right]$ & 7.281e-02 & 2.110e-02 & 2.238e-01 & 1.337e-02 \\
    $\sum \ell^2(u^\mathrm{rel}_x)$ & 1.904e+00 & 4.662e-01 & 6.606e+00 & 2.917e-01 \\
    \end{tabular}
    \caption{Average and total $\ell^2$ norm calculated from the difference of profiles given in Figure~\ref{fig:bfswss} and for $x > -70$ to avoid the area affected by inlet region.}
    \label{tab:bfsnormwss}
\end{table}
In terms of the wall shear stress and skin friction coefficients, the openHFDIBRANS and simpleFoam predictions were very similar to each other but also sensitive to the mesh resolution in the boundary layer region, see Figure~\ref{fig:bfswss} and Table~\ref{tab:bfsnormwss}. With a mesh having maximum $y^+$ of $70$ at the backward-facing step, the evolution of $\tau_\mathrm{w}$ and $C_\mathrm{f}$ mostly coincided for openHFDIBRANS and simpleFoam, see Figure~\ref{fig:bfswss}a. There was a discrepancy near the inlet, where openHFDIBRANS underestimated the peak wall shear stress. Because this effect did not propagate downstream, the most likely cause of this is that the immersed boundary representing the step started directly at the inlet leading to a slightly different boundary state than that in simpleFoam. Creating an identical inlet transition would require adding an additional domain segment with a jump down as is done in domains in Figures~\ref{fig:smeshes} and~\ref{fig:meshes3D}. However, such an addition would be against the benchmark description given in~\citep{nasa} and~\cite{driver1985}, so we accepted the presented version as the best trade-off.

\paragraph{\revs{Comparison with experimental data and other IB solvers}}
When the reattachment length was calculated from the location of zero skin friction ($x_\mathrm{reat}$) as $\ell_\mathrm{reat} = (x_\mathrm{reat}-\ell_\mathrm{step})/h_\mathrm{step}$, the solvers produced closely matching values of $4.43$ for openHFDIBRANS and $4.46$ for simpleFoam. While both values deviate considerably from the experimental result of $6.26 \pm 0.10$~\citep{driver1985}, their close agreement suggests that the discrepancy originates primarily from insufficient mesh resolution rather than from the immersed-boundary formulation. Refining the mesh near the step to achieve a maximum $y^+$ of $15$, see Figure~\ref{fig:bfswss}b, substantially improved the prediction. Although similar issues remained near the inlet, the predicted reattachment length increased to $6.08$ for openHFDIBRANS and $6.06$ for simpleFoam, both in close agreement with the measurements of~\citet{driver1985}. \revs{The effect of mesh refinement on the simulation results is studied in more detail in~\ref{app:meshref}, where both solvers showed similar convergence rates.} Overall, for both the coarse and fine meshes, openHFDIBRANS performed consistently with the standard body-fitted solver. This consistency is particularly important for future optimization studies, where reliable solver behavior is essential.

A similar level of accuracy in predicting the reattachment length was reported by~\citet{zhou2017}, who obtained a value of $6.1$. Their method is also coupled with the $k$-$\omega$ SST turbulence model and wall functions, but it uses an entirely different solution strategy designed for transient compressible flows. A transient approach is less suitable for optimization than the steady-state approach utilized by openHFDIBRANS. Still, despite targeting different flow applications, openHFDIBRANS achieved comparable accuracy to this state-of-the-art immersed boundary method.

\begin{figure}[htbp]
    \centering
    \includegraphics{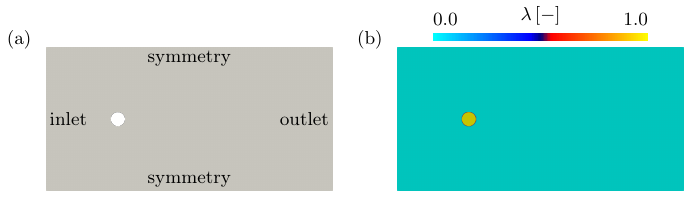}
    \caption{Flow over a smooth cylinder. (a) Domain used by simpleFoam with indicated boundaries. (b) Domain and $\lambda$ field used by openHFDIBRANS.}
    \label{fig:cyldoms}
\end{figure}
\subsection{Drag prediction for a smooth cylinder in flow}
\label{sub:cylinder}
For a smooth cylinder in flow, openHFDIBRANS predicted drag coefficients with similar  accuracy to simpleFoam, obtaining consistent values across majority of flow Reynolds numbers and mesh resolutions.

\paragraph{Simulation setup}
\label{par:simsetcyl}
The simulation domains for both solvers were made equivalent with the cylinder represented by an immersed boundary for openHFDIBRANS or mesh conformation for simpleFoam, see Figure~\ref{fig:cyldoms}. The inlet and outlet boundary conditions were the same as those in the pipe flow tests, see Section~\ref{sec:straight}. The fluid kinematic viscosity was $\nu = 10^{-6}\ \mathrm{m}^2\,\mathrm{s}^{-1}$, and the $k$-$\omega$ turbulence model was used.

The flow simulation was evaluated for Reynolds numbers ranging from $10^{-1}$ to $10^7$ where $\mathrm{Re} = \|\bm{u}_\mathrm{in}\|\,2\,r/\nu$, with $\bm{u}_\mathrm{in}$ being the inlet velocity and $r$ the radius of the cylinder. To test the dependency of results on mesh resolution, each simulation was run three times using meshes with approximately $80$, $180$ and $320$ thousand cells. From the simulation results, the drag coefficient was calculated as
\begin{equation}
\label{eq:dragcoeff}
\begin{array}{*1{>{\displaystyle}c}}
    C_\mathrm{d} = \frac{2}{\rho\,A_\mathrm{ref}\,\|\bm{u}_\mathrm{in}\|^2} \int_{\Gamma_\mathrm{sf}} \revs{\sigma_\mathrm{tot}}\cdot\bm{v}_\mathrm{d}\,\mathrm{d}S, \\ [0.4cm]
    A_\mathrm{ref} = 2\,r\,\Delta z,\ \bm{v}_\mathrm{d} = (1, 0, 0)^\mathrm{T},\ \revs{\sigma_\mathrm{tot}} = \revs{\sigma_p} + \revs{\sigma_\tau}, \\ [0.1cm] 
    \revs{\sigma_p} = -\rho\,\bm{n}\,(\tilde{p} - \tilde{p}_\mathrm{ref}),\ \revs{\sigma_\tau} = \rho\,\bm{\tau}_\mathrm{w} ,\ \bm{\tau}_\mathrm{w} = \left(\nu+\nu_\mathrm{t}\right) \left[\bm{n} \cdot \left(\nabla\bm{u} + \nabla\bm{u}^\mathrm{T}\right)_\mathrm{w}\right]\,,
\end{array}
\end{equation}
where $\rho$ represents the fluid density, \revs{$\sigma_\mathrm{tot}$ the total surface traction consisting of a pressure component $\sigma_p$ and viscous component $\sigma_\tau$}, $\Delta z$ the thickness of the computational domain in the out-of-plane direction, $\bm{n}$ the \revs{outer} surface normal, and $\tilde{p}_\mathrm{ref}$ a reference kinematic pressure. Next, $\Gamma_\mathrm{sf}$ represents the solid-fluid interface, which is formed in simpleFoam by cell faces and in openHFDIBRANS by the immersed boundary.

Lastly, this test case was utilized to study the convergence of openHFDIBRANS including the evolution of the discretization matrices condition number and solenoidality of the resulting velocity fields. The results are provided in~\ref{app:stiffandconv}\revs{, where, for low $\mathrm{Re}$, simpleFoam outperformed openHFDIBRANS in terms of convergence, but for high $\mathrm{Re}$, the convergence of both solvers was almost identical. In addition, the evolution of the condition numbers of FVM-originating system matrices was similar between the solvers, and openHFDIBRANS did not exhibit increased error in the solenoidality of the velocity field.}

\begin{figure}[htbp]
    \centering
    \includegraphics{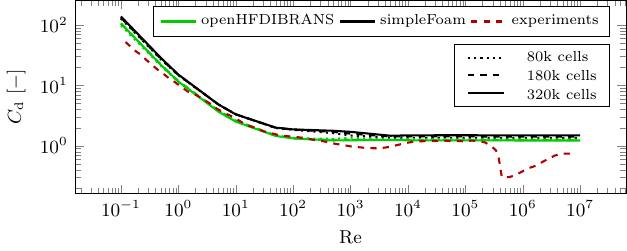}
    \caption{Evolution of the drag coefficient $C_\mathrm{d}$ for a smooth cylinder in flow.}
    \label{fig:cyldrag}
\end{figure}

\paragraph{\revs{Comparison with simpleFoam}}
\label{par:resdragcyl}
For the majority of $\mathrm{Re}$ values, openHFDIBRANS provided drag coefficients close to simpleFoam, see Figure~\ref{fig:cyldrag}. The largest discrepancies were observed near $\mathrm{Re} = 100$. In this region, the flow transitions towards the laminar von Karman vortex street, characterized by large and unsteady coherent structures. Thus, the Reynolds-averaged results are merely a crude approximation. Outside of this region, openHFDIBRANS predicted slightly lower $C_\mathrm{d}$ values, but the trends remained consistent with simpleFoam, as well as across the various mesh resolutions.

\paragraph{\revs{Comparison with experimental data and other IB solvers}}
Moreover, up to $\mathrm{Re} \approx 10^5$, the drag values predicted by the solvers were close to the compilation of experiments from~\citet{panton2013}. However, in higher $\mathrm{Re}$, neither solver was able to reproduce the drag crisis. This is probably a general limitation of RAS turbulence models. \citet{athkuri2023} showed that more advanced two-layer models are required to capture the boundary layer transitions, or transient LES or URANS methods have to be employed~\citep{ong2009}. Overall, though, openHFDIBRANS behavior was comparable with the standard mesh-conforming solver.

When compared with other state-of-the-art immersed boundary methods, openHFDIBRANS achieved similar accuracy in both low $\mathrm{Re}$ and high $\mathrm{Re}$ regions. For $\mathrm{Re} \approx 200$,~\citet{ye2020} reported drag coefficient values around $1.35$ for an incompressible and transient immersed boundary method. This value is very close to $1.33$ given by openHFDIBRANS. Using another transient immersed boundary method,~\citet{yu2025} examined high $\mathrm{Re}$ flows. However, for the smooth-cylinder case, they reported results only in the drag-crisis region, with the primary aim of assessing the performance of their URANS method. On the other hand, for a rough cylinder,~\citet{yu2025} reported data for $\mathrm{Re} \in (10^4, 10^5)$ having $C_\mathrm{d} \in (1.0,1.1)$. This is slightly lower than $1.2$ predicted by openHFDIBRANS, but as suggested by experimental studies, e.g.~\citep{zhou2015}, the drag for rough cylinders is expected to be lower than for smooth cylinders in this $\mathrm{Re}$ region. In comparison with both transient methods, openHFDIBRANS demonstrated a comparable level of accuracy. Moreover, as a steady-state formulation, it avoids time integration and may therefore offer computational advantages for flows that admit a steady-state solution.

\begin{figure}[htbp]
    \centering
    \includegraphics{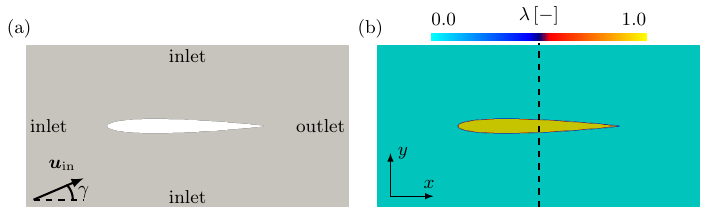}
    \caption{Flow over a NACA-0009 airfoil. (a) Computational domain used by simpleFoam with indicated boundaries and visualization of the relation between the inlet velocity $\bm{u}_\mathrm{in}$ and the angle of attack $\gamma$. (b) Domain and $\lambda$ field used by openHFDIBRANS. Black highlights the data sampling line.}
    \label{fig:nacadoms}
\end{figure}
\subsection{Lift prediction for a NACA-0009 airfoil}
\label{sub:naca}
In simulations of flow over a NACA-0009 airfoil, openHFDIBRANS accurately captured the trend in lift coefficient evolution with respect to angle of attack, although it tended to overestimate the wake size.

\paragraph{Simulation setup}
\label{par:simsetnaca}
Equivalent computational domains were used, with the NACA-0009 represented by an immersed boundary for openHFDIBRANS or mesh-conformation for simpleFoam, see Figure~\ref{fig:nacadoms}. The simulation was two-dimensional with meshes having $\sim\!10^{5}$ cells.

At the inlet boundaries, the velocity and turbulence variables were prescribed via Dirichlet boundary conditions. For pressure, the zero-gradient boundary condition was used. The inlet values of the turbulence variables were calculated according to~\citet{spalart2007}. At the outlet, the zero-gradient boundary condition for velocity and turbulence variables, and the Dirichlet boundary condition for pressure were used. The flow was simulated as incompressible with fluid kinematic viscosity $\nu \approx 10^{-6}\ \mathrm{m}^2\,\mathrm{s}^{-1}$ and the $k$-$\omega$ turbulence model was used. Three Reynolds numbers were considered, $\mathrm{Re} \in \{2\cdot10^5,\, 2\cdot10^{6},\, 2\cdot10^7\}$, where $\mathrm{Re}$ was defined with respect to the airfoil chord length. To provide an additional solver-independent reference, all investigated cases were also analyzed using XFOIL~\citep{drela1989}. The simulations were performed assuming fully turbulent flow to obtain results comparable with the RANS solutions.

For each $\mathrm{Re}$, the angle of attack was varied within the range $\gamma \in (-8,8)^\circ$ by rotating the inlet velocity vector $\bm{u}_\mathrm{in}$. The lift coefficient $C_\mathrm{\ell}$ was then calculated as
\begin{equation}
\label{eq:dlcoeffs}
\begin{array}{*1{>{\displaystyle}c}}
    C_\mathrm{\ell} = \frac{2}{\rho\,A_\mathrm{ref}\,\|\bm{u}_\mathrm{in}\|^2} \int_{\Gamma_\mathrm{sf}} \revs{\sigma_\mathrm{tot}}\cdot\bm{v}_\mathrm{\ell}\,\mathrm{d}S,\quad \bm{v}_\mathrm{\ell} = (-\sin(\gamma), \cos(\gamma),0)^\mathrm{T}\,,
\end{array}
\end{equation}
where the calculation of the total \revs{surface traction $\sigma_\mathrm{tot}$} is given in~\eqref{eq:dragcoeff}. The reference area was calculated as $A_\mathrm{ref} = \ell_\mathrm{c}\,\Delta z$ where $\ell_\mathrm{c}$ is the chord length and $\Delta z$ the thickness of the computational domain in the out-of-plane direction.

\begin{figure}[htbp]
    \centering
    \includegraphics{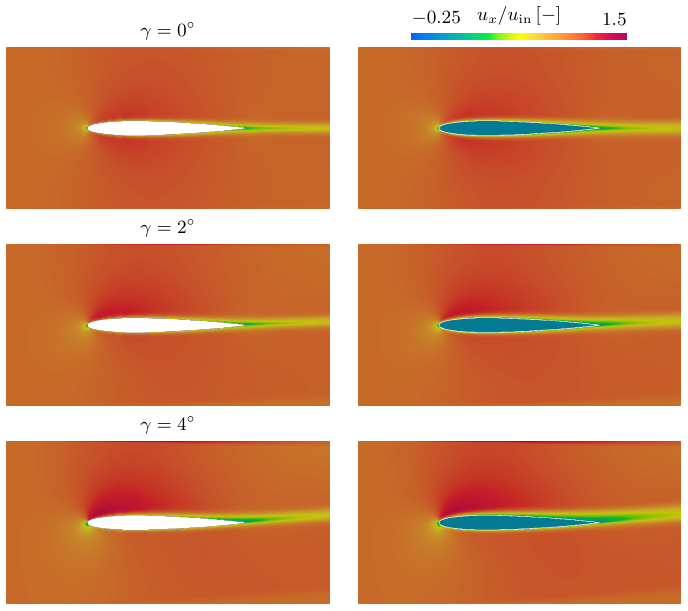}
    \caption{Simulation of flow over a NACA-0009 airfoil. Results from simpleFoam (left) and openHFDIBRANS (right) for different angles of attack $\gamma$.}
    \label{fig:nacaqua}
\end{figure}

\begin{figure}[htbp]
    \centering
    \includegraphics{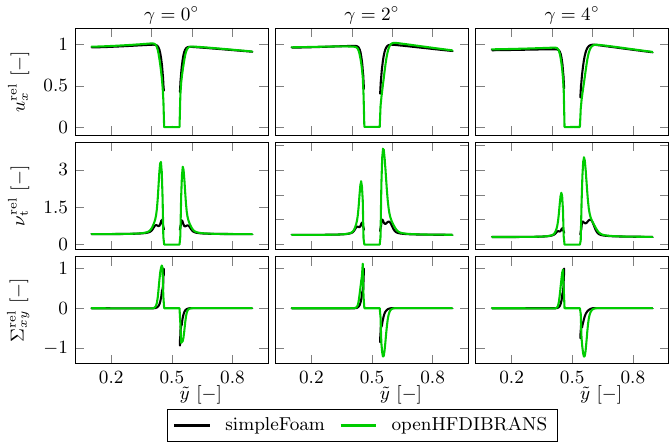}
    \caption{Field profiles from simulation of flow over a NACA-0009 airfoil with different angles of attack $\gamma$. Data were sampled along the line given in Figure~\ref{fig:nacadoms}b and normalized as in Figure~\ref{fig:straight}.}
    \label{fig:naca}
\end{figure}

\begin{table}[htbp]
    \centering
    \small
    \begin{tabular}{c|ccc}
    $\gamma$ & $0^{\circ}$ & $2^{\circ}$ & $4^{\circ}$ \\
    \toprule
    $\text{avg}\!\left[\ell^2(u^\mathrm{rel}_x)\right]$ & 3.220e-02 & 4.811e-02 & 4.786e-02 \\
    $\sum \ell^2(u^\mathrm{rel}_x)$ & 7.314e-01 & 1.093e+00 & 1.087e+00 \\
    $\text{avg}\!\left[\ell^2(\nu^\mathrm{rel}_\mathrm{t})\right]$ & 4.253e-01 & 4.606e-01 & 4.029e-01 \\
    $\sum \ell^2(\nu^\mathrm{rel}_\mathrm{t})$ & 9.661e+00 & 1.046e+01 & 9.151e+00 \\
    $\text{avg}\!\left[\ell^2(\Sigma^\mathrm{rel}_\mathrm{xy})\right]$ & 1.270e-01 & 1.529e-01 & 1.550e-01 \\
    $\sum \ell^2(\Sigma^\mathrm{rel}_\mathrm{xy})$ & 2.885e+00 & 3.474e+00 & 3.521e+00 \\
    \end{tabular}
    \caption{Average and total $\ell^2$ norm calculated from the difference of profiles given in Figure~\ref{fig:naca}.}
    \label{tab:nacanorm}
\end{table}

\begin{table}[htbp]
    \centering
    \small
    \revs{
    \begin{tabular}{c|ccc}
    $\gamma\ \mathrm{[^\circ]}$ & openHFDIBRANS $[\mathrm{m}]$ & simpleFoam $[\mathrm{m}]$ & values ratio $[-]$ \\
    \toprule
    $0$ & 0.116 & 0.095 & 1.22 \\
    $2$ & 0.117 & 0.099 & 1.18 \\
    $4$ & 0.124 & 0.105 & 1.18 \\
    $6$ & 0.148 & 0.120 & 1.23 \\
    $8$ & 0.184 & 0.148 & 1.24
    \end{tabular}
    }
    \caption{\revs{Average width of the wake behind the NACA-0009 airfoil for different angles of attack $\gamma$.}}
    \label{tab:wakewidth}
\end{table}

\paragraph{\revs{Comparison with simpleFoam}}
\label{par:resnacalift}
In a qualitative comparison, openHFDIBRANS reproduced well the dominant flow structures around the airfoil, see Figure~\ref{fig:nacaqua}, and maintained good agreement with simpleFoam in sampled field profiles, see Figure~\ref{fig:naca} and Table~\ref{tab:nacanorm}. However, the size of the wake region behind the airfoil was consistently overestimated by openHFDIBRANS. \revs{To quantify the error, the wake width was estimated as the difference between the $y$ positions above and below the airfoil where the velocity magnitude reaches $0.9\cdot\|\bm{u}_\mathrm{in}\|$. The wake estimation was done at several $x$ locations behind the airfoil ranging from $1.1\cdot\ell_\mathrm{c}$ to $1.5\cdot\ell_\mathrm{c}$, and then averaged, see Table~\ref{tab:wakewidth}. Overall, the error in wake width was about $20\,\%$ when compared with simpleFoam.}

\begin{figure}[htbp]
    \centering
    \includegraphics{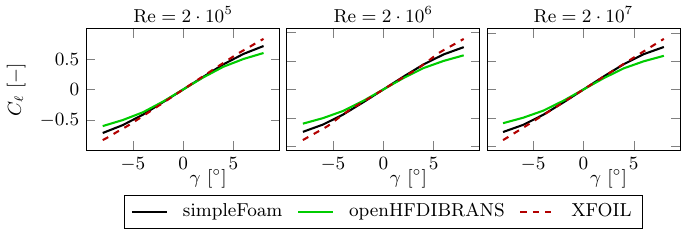}
    \caption{Evolution of the lift coefficient for NACA-0009 with respect to the angle of attack and three different $\mathrm{Re}$.}
    \label{fig:nacacoeffs}
\end{figure}

\paragraph{\revs{Comparison with reference data and other IB solvers}}
Despite the larger wake, openHFDIBRANS captured well the trends in evolution of $C_\mathrm{\ell}$ with respect to the angle of attack $\gamma$ for all investigated $\mathrm{Re}$, see Figure~\ref{fig:nacacoeffs}. For low magnitudes of $\gamma$, both openHFDIBRANS and simpleFoam agreed with $C_\mathrm{\ell}$ predicted by XFOIL. For increasing $\gamma$, the solvers slowly deviated from the XFOIL data, but openHFDIBRANS deviated more rapidly, which reflects its tendency to overestimate the wake size. Aside from the influence of the enlarged wake, the overall trend was reproduced accurately.

A similar performance of immersed boundary-RANS approaches in simulations of airfoils has been observed across several studies. In compressible simulations of high Mach number flows, \citet{zhou2017} reported that for large angles of attack, noticeable discrepancies up to $65\,\%$ in lift coefficient predictions were present between the immersed boundary method and a body-fitted approach. Similarly, the investigation of~\citet{constant2021}, where several wall modeling approaches and mesh resolutions were considered, indicated that even with a relatively fine resolution ($\sim\!10^7$ cells for a 2D NACA-0012 airfoil and $\mathrm{Re} = 6\cdot10^6$), a slight overestimation of the wake size persisted, showing that the grid refinement alone is insufficient to fully recover the fidelity of the boundary layer. Consequently, as concluded by~\citet{troldborg2022}, the development of new IB-adjusted wall models appears necessary, particularly for a more accurate inclusion of the boundary layer transition. Nevertheless, the present results demonstrate that openHFDIBRANS performs consistently with other state-of-the-art immersed boundary methods.

\begin{figure}[htbp]
    \centering
    \includegraphics{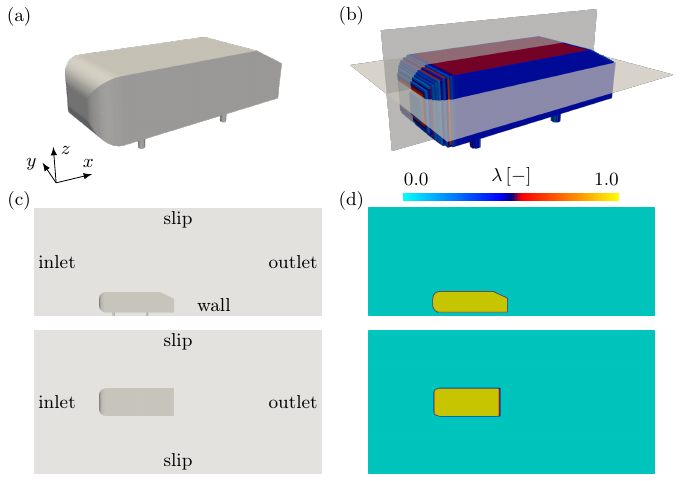}
    \caption{(a) Visualization of the Ahmed body. (b) Representation of the Ahmed body using the $\lambda$ field. (c) Vertical and horizontal view on the Ahmed body inside a computational domain with indicated boundaries. (d) Vertical and horizontal slice of the $\lambda$ field.}
    \label{fig:ahmeddom}
\end{figure}
\subsection{Recirculation region behind an Ahmed body}
\label{sub:ahmed}
In the the most complex benchmark case, openHFDIBRANS reproduced well the characteristics of the recirculation region, but shares a common limitation of immersed boundaries present in the representation of sharp edges.

\paragraph{Simulation setup}
\label{par:simsetahmed}
In this benchmark, we followed the experimental conditions described in~\citet{lienhart2003}. The Ahmed body with slant angle $25^\circ$ was positioned inside a block domain where the body itself was represented by immersed boundary for openHFDIBRANS and mesh conformation for simpleFoam, see Figure~\ref{fig:ahmeddom}. The computational mesh comprised $\sim 2\cdot10^{7}$ cells.

At the domain inlet and outlet, the same boundary conditions as for pipe flow were prescribed, see Section~\ref{sec:straight}. Next, the boundary below the Ahmed body was treated as a wall with the no-slip condition for velocity, and zero-gradient condition for the other variables. At the remaining boundaries, the slip boundary condition was used for velocity, and zero-gradient for the rest. The fluid kinematic viscosity was $\nu = 1.5\cdot10^{-5}\ \mathrm{m}^2\,\mathrm{s}^{-1}$ and the $k$-$\omega$ model was used for turbulence closure.

Two Reynolds number regimes were considered, low regime with $\mathrm{Re} = 7.68\cdot10^3$ and high regime with $\mathrm{Re} = 7.68\cdot10^5$, where $\mathrm{Re}$ is calculated with respect to the Ahmed body height $h_\mathrm{body} = 0.288\ \mathrm{m}$. For the high $\mathrm{Re}$ regime, experimental data are available in~\citep{lienhart2003}. The low $\mathrm{Re}$ regime was included to test the consistency of openHFDIBRANS behavior.

Given the computational cost associated with the considered mesh size, the case was computed in parallel; a detailed study of the parallel scaling of openHFDIBRANS, including comparison with simpleFoam, is presented in~\ref{app:parallel}. The study showed that openHFDIBRANS scales similarly to simpleFoam and \revs{introduces minimal computational overhead.}

\begin{figure}[htbp]
    \centering
    \includegraphics{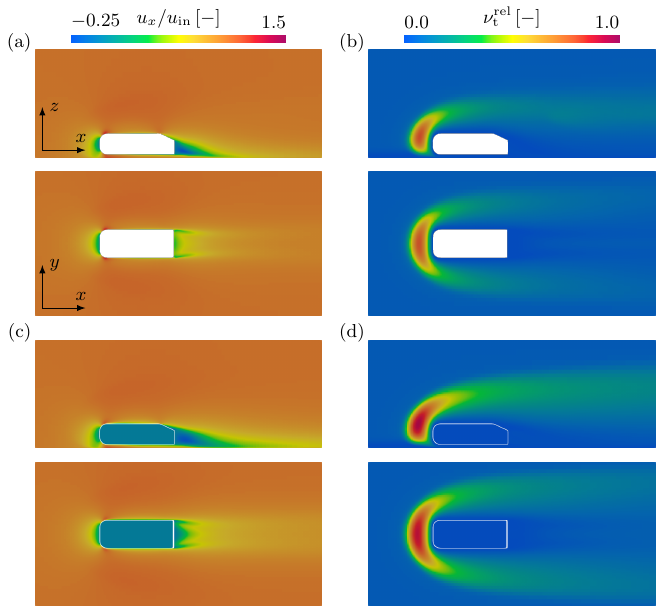}
    \caption{Ahmed body simulation with $\mathrm{Re} = 7.68\cdot10^3$. Vertical and horizontal slices through flow fields from the simulation of flow over an Ahmed body. (a-b) Fields from simpleFoam. (c-d) Fields from openHFDIBRANS.}
    \label{fig:ahmedfieldslow}
\end{figure}

\begin{figure}[htbp]
    \centering
    \includegraphics{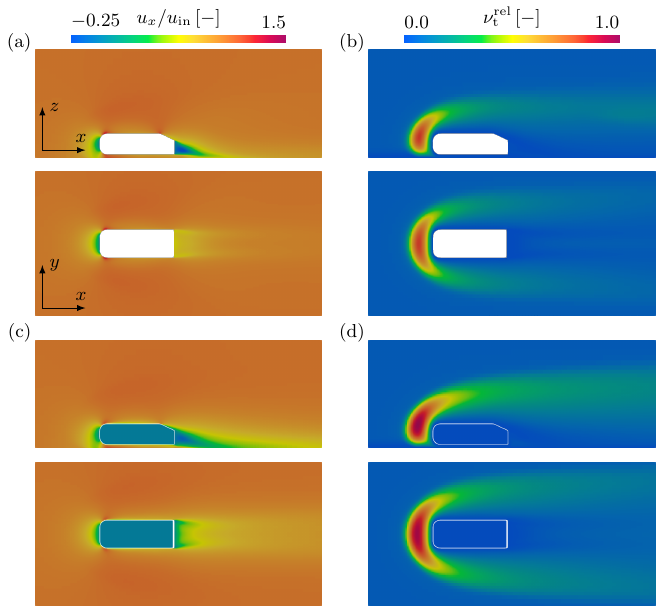}
    \caption{Ahmed body simulation with $\mathrm{Re} = 7.68\cdot10^5$. Vertical and horizontal slices through flow fields from the simulation of flow over an Ahmed body. (a-b) Fields from simpleFoam. (c-d) Fields from openHFDIBRANS.}
    \label{fig:ahmedfieldshigh}
\end{figure}

\begin{figure}[htbp]
    \centering
    \includegraphics{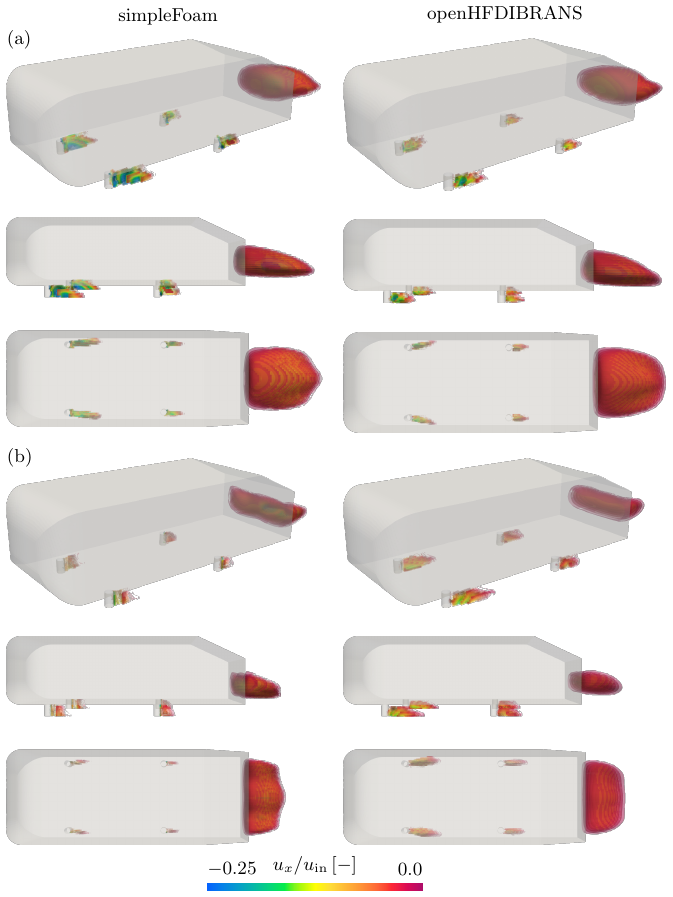}
    \caption{Different views onto the recirculation regions behind the Ahmed body for (a) $\mathrm{Re} = 7.68\cdot10^3$ and (b) $\mathrm{Re} = 7.68\cdot10^5$. Visualized as contours of the velocity field.}
    \label{fig:ahmedrecirc}
\end{figure}

\paragraph{\revs{Comparison with simpleFoam}}
\label{par:resahmed}
A qualitative comparison of the $u_x$ and $\nu_\mathrm{t}$ fields showed that openHFDIBRANS accurately reproduced the dominant flow structures in both low $\mathrm{Re}$, see Figure~\ref{fig:ahmedfieldslow}, and high $\mathrm{Re}$, see Figure~\ref{fig:ahmedfieldshigh}, regimes. However, similarly to the test scenario with NACA-0009, the openHFDIBRANS solver mostly overestimated the size of the recirculation region, see velocity-contour visualization of the recirculation regions in Figure~\ref{fig:ahmedrecirc}.

\begin{table}[htbp]
    \centering
    \small
    \begin{tabular}{c|cccc}
    low $\mathrm{Re}$ & $\ell_\mathrm{rec}\ \mathrm{[m]}$ & \revs{values} ratio $[-]$ & $V_\mathrm{rec}\ \mathrm{[m^3]}$ & \revs{values} ratio $[-]$ \\
    \toprule
    openHFDIBRANS & $0.356$ & $0.94$ & $0.0094$ & $1.03$ \\
    simpleFoam & $0.377$ & -- & $0.0091$ & -- \\
    & & & & \\
    high $\mathrm{Re}$ & $\ell_\mathrm{rec}\ \mathrm{[m]}$ & ratio $[-]$ & $V_\mathrm{rec}\ \mathrm{[m^3]}$ & ratio $[-]$ \\
    \toprule
    openHFDIBRANS & $0.230$ & $1.10$ & $0.0055$ & $1.04$ \\
    simpleFoam & $0.209$ & -- & $0.0053$ & -- \\
    \end{tabular}
    \caption{Characteristic length and volume of the recirculation region behind an Ahmed body.}
    \label{tab:ahmedrecirchar}
\end{table}

For both low and high $\mathrm{Re}$, the openHFDIBRANS estimates of the recirculation region characteristic length $\ell_\mathrm{rec}$ and volume $V_\mathrm{rec}$ were mostly slightly larger than those of simpleFoam, see Table~\ref{tab:ahmedrecirchar}. To calculate the characteristics, computational cells with negative $u_x$ were extracted. Then, the length of the recirculation region was estimated as the distance from the Ahmed body rear end and the center of the furthest cell. The volume was calculated as a sum of the individual cell volumes located behind the body. Still, for both $\mathrm{Re}$ regimes, the characteristic values predicted by openHFDIBRANS were within $10\,\%$ of simpleFoam with the biggest difference in the length of the recirculation region for high $\mathrm{Re}$ regime. 

\begin{figure}[htbp]
    \centering
    \includegraphics{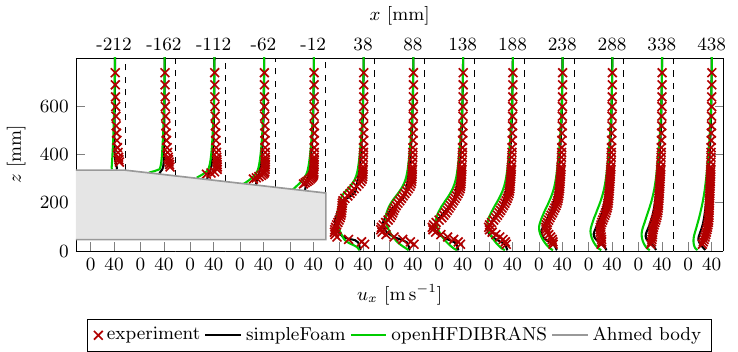}
    \caption{Flow over an Ahmed body with $\mathrm{Re} = 7.68\cdot10^5$. Profiles of $u_x$ were sampled on the $x$-$z$ plane defined by $y = 0.0$.}
    \label{fig:ahmedprofiles}
\end{figure}

\paragraph{\revs{Comparison with experimental data and other IB solvers}}
\revs{When compared with experimental data of~\citet{lienhart2003}, the overprediction of recirculation region size} is \revs{also} visible \revs{in} the comparison of $u_x$ profiles, see most-right profiles in Figure~\ref{fig:ahmedprofiles}. \revs{More specifically, in~\citet{lienhart2003}, it is stated that the recirculation region disappeared at $x = 200\,\mathrm{mm}$. With respect to this value, both solvers overshot the recirculation region length, simpleFoam by $4.5\,\%$ and openHFDIBRANS by $15\,\%$.} Comparable results from Ahmed body simulation with an immersed boundary were reported by~\citet{onishi2021}. Even though their immersed boundary method was coupled with LES models, their velocity profiles computed on coarser grids showed the same tendency to overestimate the recirculation region length as our results. When looking at other (U)RANS methods, a similar level of agreement between a body-fitted mesh and immersed boundary in flow separation behind a bluff body has been reported, for example, by~\citet{constant2024}. A likely reason is the limited geometric resolution of the immersed boundary near the sharp edges of the Ahmed body when using the relatively coarse meshes typically employed in RANS simulations. Simultaneously, the observed differences in the recirculation zone structure are within the range reported by~\citet{delassaux2021} for small geometric variations of the edges of the Ahmed body slanted surface. Despite the limitations associated with sharp edges, openHFDIBRANS successfully reproduced the flow structures for this complex geometry.

\subsection{Prediction of coolant flow and retention time in a die-casting die core}
\label{sub:cooling}
In this \revs{simulation example motivated by an} engineering application focused on coolant flow in a die-casting die core, openHFDIBRANS predicted both velocity fields and coolant retention times with accuracy comparable to simpleFoam.

\begin{figure}[htbp]
    \centering
    \includegraphics{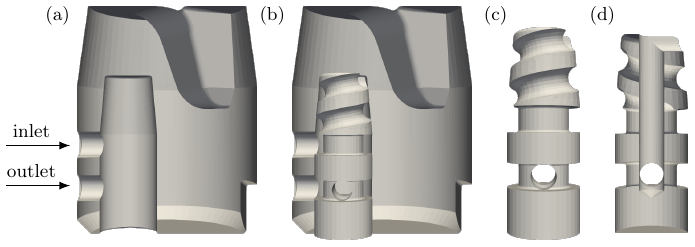}
    \caption{Visualization of the die-casting die. (a) Clip of the die geometry with indicated inlet and outlet to the cooling cavity. (b) Die with the flow guiding insert. (c) Geometry of the insert. (d) Clip of the insert geometry.}
    \label{fig:coolinggeometries}
\end{figure}

\paragraph{Simulation setup}
\label{par:simsetcooling}
The die-casting die core operates in a thermal cycle where it repeatedly comes into contact with molten metal injected at pressures up to tens of $\mathrm{MPa}$~\citep{murray2011}, and then it has to cool down before it can be used again. Thus, the cooling rate directly affects the manufacturing efficiency. Cooling is achieved by circulating a liquid coolant through an internal cavity inside the die, see Figure~\ref{fig:coolinggeometries}a. To improve cooling efficiency, a flow-guiding insert is placed in the cavity, see Figure~\ref{fig:coolinggeometries}b-d.

\begin{figure}[htbp]
    \centering
    \includegraphics{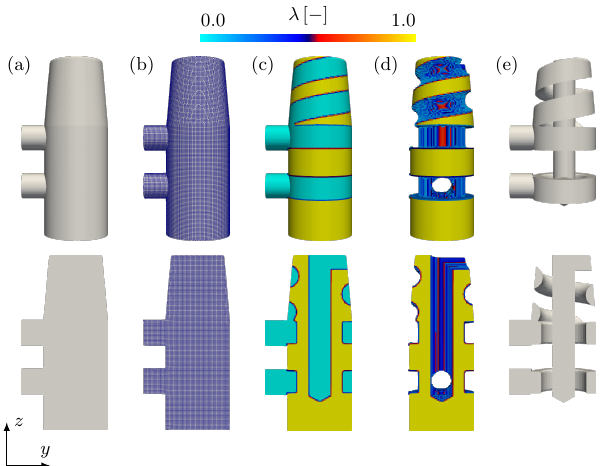}
    \caption{Side view (top) and clip along the $y$-$z$ plane through the middle of (a) the domain representing the cooling cavity used for openHFDIBRANS, (b) mesh in the cavity domain, (c) $\lambda$ field representing the insert, (d) cells of the domain with $\lambda > 0.0$, and (e) domain used for simpleFoam.}
    \label{fig:coolingdomains}
\end{figure}
To simulate the coolant flow inside the cavity, equivalent domains were created for openHFDIBRANS and simpleFoam, see Figure~\ref{fig:coolingdomains}. For openHFDIBRANS, the geometry of the insert was represented by an immersed boundary, while for simpleFoam the mesh was fitted to the geometry. Both domains included only an inlet, outlet and wall boundaries. At each boundary, the boundary conditions were prescribed in the same manner as in the 2D pipe flow tests, see Section~\ref{sec:straight}. The fluid kinematic viscosity was $\nu = 10^{-6}\,\mathrm{m}^{2}\,\mathrm{s}^{-1}$ and the inlet flow Reynolds number was $\mathrm{Re} = 11\,600$. The resolution of the computational meshes was selected to ensure $y^+ \lesssim 250$. 

\revs{The distribution of $y^+$ values encountered during the simulation was nearly identical for simpleFoam and openHFDIBRANS, and for the latter, it is given in Table~\ref{tab:yPlusDistributionInPracticalApplication}. Both simulations were computed with the switching treatment of wall modeling, see~\eqref{eq:omegaws} and~\eqref{eq:alphauk}. In practical applications, the $y^+$ field will generally span the buffer layer and the employed switching treatment therefore represents an approximation. In the present case, a majority of the cells (approximately $60\,\%$) were located either inside the viscous sublayer or in the logarithmic region. Nevertheless, satisfactory results were obtained despite approximately $40\,\%$ of the wall-adjacent cells residing in the buffer layer. These results suggest that the switching treatment remains practically applicable for moderately broad $y^+$ distributions, although the interval $5<y^+<11$ remains the most sensitive owing to the transition between the two wall-treatment regimes.}
\begin{table}[htbp]
    \centering
    \revs{
    \begin{tabular}{l|cccc}
         \toprule
         $y^+$        ($-$)   & $<5$ & $5-11$ & $11-30$ & $>30$ \\
         cell fraction ($\%$) & 10   & 13     & 27      & 50    \\ 
         \midrule
         true region          & viscous & \multicolumn{2}{c}{buffer} & logarithmic \\
         treated as           & \multicolumn{2}{c}{viscous} & \multicolumn{2}{c}{logarithmic}\\
         \bottomrule
    \end{tabular}
    \caption{\revs{Distribution of $y^+$ values in openHFDIBRANS simulation of the cooling cavity.}}
    \label{tab:yPlusDistributionInPracticalApplication}
    }
\end{table}

\begin{figure}[htpb]
    \centering
    \includegraphics{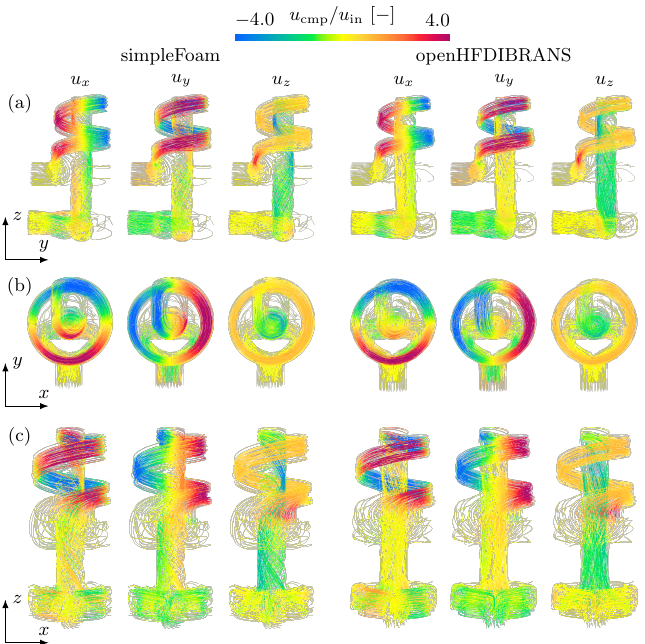}
    \caption{(a) Side, (b) top, and (c) front view onto the velocity streamlines in the coolant flow. Each view is drawn separately for each velocity component (cmp).}
    \label{fig:coolingstreamlines}
\end{figure}

\begin{figure}[htbp]
    \centering
    \includegraphics{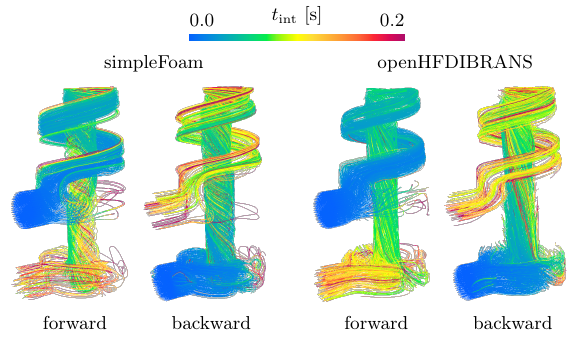}
    \caption{Visualization of streamlines colored by the integration time $t_\mathrm{int}$. For each simulation, two types of streamlines are shown. Left are those obtained by forward integration from the inlet and right those obtained by backward integration starting from the outlet.}
    \label{fig:coolinginttime}
\end{figure}
\paragraph{Comparison of velocity streamlines and retention times}
\label{par:coolingveltimes}
Our openHFDIBRANS solver reproduced well the main flow structures inside the 3D geometry and showed good agreement with simpleFoam across all velocity components, see streamlines in Figure~\ref{fig:coolingstreamlines}. The most noticeable difference was the flow behavior inside the middle vertical pipe, where the fluid transfers to the bottom of the cooling cavity and exits. There, the simpleFoam solution appeared to retain more angular momentum because the streamlines formed more loops before reaching the bottom.

\begin{figure}[htbp]
    \centering
    \includegraphics{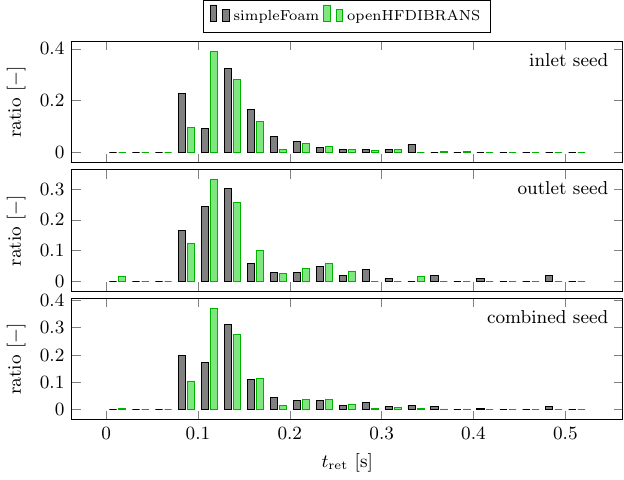}
    \caption{Distribution of the retention times $t_\mathrm{ret}$ estimated from streamlines seeded at the inlet, outlet, or both combined.}
    \label{fig:coolinghist}
\end{figure}

\begin{table}[htbp]
    \centering
    \small
    \begin{tabular}{c|cccc}
    average $t_\mathrm{ret}\ \mathrm{[s]}$ & inlet seed & outlet seed & combined seed \\
    \toprule
    openHFDIBRANS & $0.139$ & $0.141$ & $0.139$ \\
    simpleFoam & $0.147$ & $0.156$ & $0.152$ \\
    & & & & \\
    median $t_\mathrm{ret}\ \mathrm{[s]}$ & inlet seed & outlet seed & combined seed \\
    \toprule
    openHFDIBRANS & $0.126$ & $0.128$ & $0.127$ \\
    simpleFoam & $0.138$ & $0.132$ & $0.134$ \\
    \end{tabular}
    \caption{Average and median retention time calculated from distributions given in Figure~\ref{fig:coolinghist}.}
    \label{tab:coolingtimes}
\end{table}

From the analysis of the retention times, both openHFDIBRANS and simpleFoam produced similar time distributions, see Figure~\ref{fig:coolinghist}. The retention time $t_\mathrm{ret}$ was estimated as the integration time required for a streamline to reach the outlet when starting at the inlet, or vice versa. For openHFDIBRANS, the distribution peak occurred at $\sim 0.12\ \mathrm{s}$ for all seeding strategies. For simpleFoam, the peaks were slightly later at $\sim 0.14\ \mathrm{s}$. This shift agrees with the difference in streamline behavior observed in the middle pipe in Figure~\ref{fig:coolinginttime}. The distribution data were further analyzed by calculating the average and median retention time, for which the openHFDIBRANS predictions were slightly lower than those of simpleFoam, see Table~\ref{tab:coolingtimes}. Apart from the slight shift in the retention time distributions, the differences between openHFDIBRANS and simpleFoam ranged from $3\,\%$ to $11\,\%$ which is satisfactory. 

\revs{To establish the predictive accuracy of both solvers, a comparison with industrial measurements would be required. Such measurements, however, focus mostly on the temperature evolution during the die cooling. Thus, coupling of openHFDIBRANS with heat transfer modeling is planned.} Nonetheless, \revs{in comparison with the standard CFD solver simpleFoam}, openHFDIBRANS preserved the key characteristics of the cooling system, supporting its future application in geometry optimization studies. 

\section{Conclusion}
\label{sec:conc}

\bigskip
In this work, an immersed-boundary solver for steady-state Reynolds-averaged Navier--Stokes (RANS) simulations has been developed by extending the hybrid fictitious domain--immersed boundary (HFDIB) methodology to include two-equation turbulence models, wall functions and steady-state solution algorithms. The resulting solver, denoted \textit{openHFDIBRANS}, is implemented as a fully open-source extension of OpenFOAM with an emphasis on robustness, reproducibility, and applicability to engineering flows involving complex geometries.

The principal findings and their significance can be summarized as follows:
\begin{itemize} 
    \item \textit{Extension of HFDIB to steady-state RANS simulations.} The proposed methodology successfully extends the HFDIB framework to the solution of incompressible steady-state RANS equations coupled with the k--$\omega$, k--$\omega$ SST, k--$\varepsilon$, and realizable k--$\varepsilon$ turbulence models. This broadens the applicability of HFDIB from particle-resolved and DNS-oriented simulations to engineering-relevant turbulent flows.
    \item \textit{Robust immersed-boundary treatment of near-wall turbulence.} The developed wall-function strategy and immersed-boundary forcing formulation provide stable coupling between momentum and turbulence transport equations. The proposed approach remains applicable across a wide range of Reynolds numbers and geometric configurations without requiring body-fitted meshes.
    \item \textit{Improved consistency between immersed-boundary forcing and incompressibility enforcement.} The modified continuity-equation formulation, in which incompressibility is enforced primarily within the fluid region, alleviates a known source of inconsistency in immersed-boundary methods and contributes to stable solver convergence.
    \item \textit{Good agreement with body-fitted reference simulations.} Across all benchmark configurations as well as in a real-life engineering application the predicted velocity fields, turbulence quantities, and integral measures showed generally good agreement with body-fitted \textit{simpleFoam} simulations and, where available, experimental reference data.
    \item \textit{Robustness with respect to numerical and geometric factors.} Additional studies demonstrated that the solver remains reasonably insensitive to variations in Reynolds number, turbulence-model selection, immersed-boundary position, mesh-interface misalignment, discretization schemes, mesh refinement, and parallel execution settings. Such robustness is particularly important for design and optimization workflows.
    \item \textit{Minimal computational overhead and preserved parallel scalability.} The proposed immersed-boundary formulation showed parallel-scaling characteristics comparable to those of the standard body-fitted solver, indicating that the additional immersed-boundary treatment does not introduce a significant performance penalty.
    \item \textit{Reproducible open-source implementation.} To the best of our knowledge, \textit{openHFDIBRANS} represents one of the few fully open-source immersed-boundary frameworks combining steady-state RANS turbulence modeling, wall functions, and comprehensive benchmark validation. The public availability of the source code, tutorial cases, and validation studies facilitates reproducibility and future methodological developments.
\end{itemize}

Although some systematic discrepancies remain, particularly a tendency to overestimate wake and recirculation-region sizes in cases \revs{where the limited ability of immersed boundary methods to fully resolve sharp edges is combined with flow separation (NACA-0009, Ahmed body)}, the method demonstrated consistent behavior across a wide range of test cases. Consequently, \textit{openHFDIBRANS} appears well suited for engineering applications where geometric flexibility, robustness, and computational efficiency are of primary importance, such as shape and topology optimization, industrial flow analysis, and multiphysics process design.

Future work will focus on the application of the solver to optimization problems in engineering systems, further refinement of immersed-boundary wall treatments, and the development of improved turbulence-modeling strategies for separated and transition-dominated flows. The complete source code and accompanying examples are publicly available at:
\begin{center} 
    \url{https://github.com/techMathGroup/openHFDIBRANS} 
\end{center}

\subsection*{Data and code availability statement}
\noindent Ongoing development is hosted in a public GitHub repository~\cite{github2026}: \url{https://github.com/techMathGroup/openHFDIBRANS}.

\noindent An archived snapshot for reproducing the results presented in this paper is available on Zenodo~\cite{kubickova2026zenodo}: \url{https://zenodo.org/records/22917049}. The archive contains:
\begin{itemize}
    \item an installation script,
    \item OpenFOAM implementation of a steady-state solver, library, and processing utilities,
    \item tutorial cases based on benchmarks presented in this paper, and
    \item example cases with simulation settings prepared for a parallel run.
\end{itemize}
\noindent The provided materials are released under the Creative Commons Attribution 4.0 International License (CC BY 4.0).

\subsection*{Credit author statement}
\textbf{Lucie Kubíčková} -- Formal analysis; Investigation; Methodology; Software; Validation; Visualization; Writing -- Original Draft. \textbf{Martin Isoz} -- Conceptualization; Formal Analysis; Funding acquisition; Investigation; Methodology; Supervision; Writing -- Review \& Editing.


\subsection*{Declaration of AI usage}
During the preparation of this work, the authors used Claude (Anthropic's AI assistant) and Microsoft Copilot to improve the language and readability of the manuscript and consulted a professional editor regarding style and presentation. The authors subsequently reviewed and edited all text and assume full responsibility for the content of the published article.

\subsection*{Acknowledgments}
{\small{
The work was financially supported by the Czech Science Foundation project 25-17815S. Also, the authors acknowledge the financial support provided by the Ministry of Education, Youth, and Sports of the Czech Republic via the project No. CZ.02.01.01/00/23\_020/0008501 (METEX), co-funded by the European Union. Besides, this work was supported by the institutional support RVO:61388998 and from the grant of Specific university research – A1\_FCHI\_2026\_004. Finally, we would like to thank the company Beneš a Lát, a.s. for motivating the real-life engineering application.
}}

\subsection*{Nomenclature}
\label{sec:nomen}

\begin{tabular}{ rl }
$A$ & area, {$[\lUnit^2]$} \\
$\bm{b}$ & vector of bounding box size, {$[\lUnit]$} \\
$C_\mathrm{d}$ & drag coefficient, {$[\ndUnit]$} \\
$C_\mathrm{\ell}$ & lift coefficient, {$[\ndUnit]$} \\
$C_\mathrm{f}$ & skin friction coefficient, {$[\ndUnit]$} \\
$d$ & diameter, {$[\lUnit]$} \\
$D$ & diagonal matrix, {$[\ndUnit]$} \\
$\bm{f}$ & momentum source term, {$[\aUnit]$} \\
$h$ & height, {$[\lUnit]$} \\
$I$ & turbulence intensity, {$[\ndUnit]$} \\
$k$ & turbulence kinetic energy, {$[\lUnit^2\,\tUnit^{-2}]$} \\
$\ell$ & length, {$[\lUnit]$} \\
$\ell^2$ & square norm, {$[\ndUnit]$} \\
$L$ & lower triangular matrix, {$[\ndUnit]$} \\
$M$ & discretization matrix of momentum equations, {$[\ndUnit]$} \\
$\mathcal{M}$ & operator of the momentum transport, {$[\ndUnit]$} \\
$\bm{n}$ & outer unit normal, {$[\ndUnit]$} \\
$\mathcal{N}$ & operator of the turbulence kinetic energy transport, {$[\ndUnit]$} \\
$\tilde{p}$ & average kinematic pressure, {$[\lUnit^2\,\tUnit^{-2}]$} \\
$P_k$ & production of turbulence kinetic energy, {$[\lUnit^2\,\tUnit^{-3}]$} \\
$r$ & radius, {$[\lUnit]$} \\
$\mathrm{Re}$ & Reynolds number, {$[\ndUnit]$} \\
$S$ & turbulence kinetic energy source term, {$[\lUnit^2\,\tUnit^{-3}]$} \\
$t$ & time, {$[\tUnit]$} \\
$\bm{u}$ & average velocity, {$[\vUnit]$} \\
$u_\tau$ & friction velocity, {$[\vUnit]$} \\
$U$ & upper triangular matrix, {$[\ndUnit]$} \\
$v$ & vertex, {$[\ndUnit]$} \\
$\bm{v}$ & direction vector, {$[\ndUnit]$} \\
$V$ & volume, {$[\lUnit^3]$} \\
$x, y, z$ & spatial coordinates, {$[\lUnit]$} \\
$y^+$ & normalized wall distance, {$[\ndUnit]$} \\
$y_\perp$ & perpendicular distance, {$[\lUnit]$} \\
\end{tabular}

\subsubsection*{Greek letters}
\begin{tabular}{ rl }
$\alpha$ & mask field, {$[\ndUnit]$} \\
$\beta$ & angle of misalignment, {$[^\circ]$} \\
$\gamma$ & angle of attack, {$[^\circ]$} \\
$\Gamma$ & interface, {$[\ndUnit]$} \\
$\delta$ & cell size, {$[\lUnit]$} \\
$\Delta$ & domain thickness, {$[\lUnit]$} \\
$\epsilon$ & tolerance, {$[\ndUnit]$} \\
$\varepsilon$ & dissipation of turbulence kinetic energy, {$[\lUnit^2\,\tUnit^{-3}]$} \\
$\kappa$ & matrix condition number, {$[\ndUnit]$} \\
$\lambda$ & solid phase indicator field, {$[\ndUnit]$} \\
$\lambda^\mathrm{e}$ & eigenvalues, {$[\ndUnit]$} \\
$\nu$ & kinematic viscosity, {$[\lUnit^2\,\tUnit^{-1}]$} \\
$\rho$ & fluid density, {$[\mUnit\,\lUnit^{-3}]$} \\
\revs{$\sigma$} & \revs{traction}, \revs{{$[\mUnit\,\lUnit^{-1}\,\tUnit^{-2}]$}} \\
$\sigma_\perp$ & signed perpendicular distance, {$[\lUnit]$} \\
$\Sigma_{xy}$ & $x$-$y$ component of the kinematic Reynolds stress tensor, {$[\lUnit^2\,\tUnit^{-2}]$} \\
$\tau_\mathrm{w}$ & wall shear stress, {$[\lUnit^2\,\tUnit^{-2}]$} \\
$\omega$ & specific dissipation of turbulence kinetic energy, {$[\tUnit^{-1}]$} \\
$\Omega$ & spatial computational domain, {$[\ndUnit]$} \\
$\Omega^h$ & discrete spatial domain, {$[\ndUnit]$} \\
$\Omega^h_P$ & computational cell, {$[\ndUnit]$} \\
\end{tabular}

\subsubsection*{Subscripts and superscripts}
\begin{tabular}{ rl}
$\mathrm{atm}$ & atmospheric \\
$\mathrm{avg}$ & average \\
$\mathrm{bc}$ & boundary cells \\
$\mathrm{c}$ & chord \\
$\mathrm{eff}$ & effective \\
$\mathrm{f}$ & fluid \\
$\mathrm{fc}$ & free-stream cells \\
$\mathrm{ib}$ & immersed boundary \\
$\mathrm{ic}$ & in-solid cells \\
$\mathrm{in}$ & inlet \\
$\mathrm{int}$ & integration \\
$\mathrm{lam}$ & laminar \\
$\mathrm{log}$ & logarithmic region or scale \\
$\mathrm{out}$ & outlet \\
$\mathrm{old}$ & previous iteration or initial guess \\
$\mathrm{reat}$ & reattachment \\
$\mathrm{ret}$ & retention \\
$\mathrm{ref}$ & reference \\
$\mathrm{rel}$ & relative \\
$\mathrm{s}$ & solid \\
$\mathrm{t}$ & turbulent \\
$\mathrm{tot}$ & total \\
$\mathrm{sf}$ & solid-fluid interface \\
$\mathrm{vis}$ & viscous sublayer \\
\end{tabular}

\subsubsection*{Abbreviations}
\begin{tabular}{ rl}
CFD & computational fluid dynamics \\
DEM & discrete element method \\
HFDIB & hybrid fictitious domain-immersed boundary method \\
IB & immersed boundary \\
RANS & Reynolds-averaged Navier-Stokes equations \\
sF & simpleFoam \\
TMR & turbulence modeling resource \\
\end{tabular}


\bibliography{references}

@article{patankar1972,
    author = {S.V Patankar and D.B Spalding},
    title = {A calculation procedure for heat, mass and momentum transfer in three-dimensional parabolic flows},
    journal = {International Journal of Heat and Mass Transfer},
    volume = {15},
    number = {10},
    pages = {1787--1806},
    year = {1972},
    issn = {0017-9310},
    doi = {10.1016/0017-9310(72)90054-3},
}

@TechReport{menter1992,
    author = {F. R. Menter},
    title = {Improved two equation $k$-$\omega$ turbulence models for aerodynamic flows},
    number = {N93-22809},
    institution = {NASA},
    year = {1992},
    pages = {34},
}

@article{launder1974,
    author = {B.E. Launder and D.B. Spalding},
    title = {The numerical computation of turbulent flows},
    journal = {Computer Methods in Applied Mechanics and Engineering},
    volume = {3},
    pages = {269--289},
    year = {1974},
    doi = {0.1016/0045-7825(74)90029-2},
}

@article{tahry1983,
    author = {S.H. El Tahry},
    title = {$k$-$\epsilon$ equation for compressible reciprocating engine flows},
    journal = {Energy},
    volume = {7},
    pages = {345--353},
    year = {1983},
    doi = {10.2514/3.48086},
}

@article{shih1995,
    author = {T. H. Shih and W.W. Liou and A. Shabbir and Z. Yang and J. Zhu},
    title = {A new $k$-$\epsilon$ eddy viscosity model for high {R}eynolds number turbulent flows},
    journal = {Computer \& Fluids},
    volume = {24},
    pages = {227--238},
    year = {1995},
    doi = {10.1016/0045-7930(94)00032-T},
}

@article{russo2016,
    author = {F. Russo and N. T. Basse},
    title = {Scaling of turbulence intensity for low-speed flow in smooth pipes},
    journal = {Flow Measurement and Instrumentation},
    year = {2016},
    volume = {52},
    pages = {101--114},
    doi = {10.1016/j.flowmeasinst.2016.09.012},
}

@article{peskin1972,
    author = {C. S. Peskin},
    title = {Flow patterns around heart valves: A numerical method},
    journal = {Journal of Computational Physics},
    volume = {10},
    number = {2},
    pages = {252--271},
    year = {1972},
    doi = {10.1016/0021-9991(72)90065-4},
}

@article{isoz2022,
    author = {M. Isoz and M. Kotouč Šourek and O. Studeník and P. Kočí},
    title = {Hybrid fictitious domain-immersed boundary solver coupled with discrete element method for simulations of flows laden with arbitrarily-shaped particles},
    journal = {Computers \& Fluids},
    volume = {244},
    pages = {105538},
    year = {2022},
    doi = {10.1016/j.compfluid.2022.105538},
}

@article{kotouc2024,
    title = {Viscosity Prediction for Dense Suspensions of Non-spherical Particles Based on {CFD}-{DEM} Simulations},
    journal = {Powder Technology},
    volume = {444},
    pages = {120067},
    year = {2024},
    doi = {10.1016/j.powtec.2024.120067},
    author = {Martin Kotouč Šourek and Ondřej Studeník and Martin Isoz and Petr Kočí and Andrew P.E. York},
}

@article{municchi2017,
    author = {F. Municchi and S. Radl},
    title = {Consistent closures for Euler-Lagrange models of bi-disperse gas-particle suspensions derived from particle-resolved direct numerical simulations},
    journal = {International Journal of Heat and Mass Transfer},
    volume = {111},
    pages = {171--190},
    year = {2017},
    doi = {10.1016/j.ijheatmasstransfer.2017.03.122},
}

@article{municchi2018,
    author = {F. Municchi and S. Radl},
    title = {Momentum, heat and mass transfer simulations of bounded dense mono-dispersed gas-particle systems},
    journal = {International Journal of Heat and Mass Transfer},
    volume = {120},
    pages = {1146--1161},
    year = {2018},
    doi = {10.1016/j.ijheatmasstransfer.2017.12.105},
}

@inproceedings{kubickova2023,
    author = {L. Kubíčková and M. Isoz},
    title = {On {R}eynolds-Averaged Turbulence Modeling with Immersed Boundary Method},
    booktitle = {Proceedings of Topical Problems of Fluid Mechanics 2023},
    editor = {D. Šimurda and T. Bodnár},
    month = {02},
    pages = {104--111},
    publisher = {IT CAS},
    year = {2023},
    doi = {10.14311/TPFM.2023.015},
}

@article{capizzano2011,
  title = {Turbulent Wall Model for Immersed Boundary Methods},
  author = {F. Capizzano},
  journal = {AIAA Journal},
  volume = {49},
  number = {11},
  pages = {2367--2381},
  year = {2011},
  doi = {10.2514/1.J050466},
}

@article{troldborg2022,
  title = {Immersed boundary method for the incompressible {R}eynolds {A}veraged {N}avier–{S}tokes equations},
  author = {N. Troldborg and N. N. Sørensen and F. Zahle},
  journal = {Computers \& Fluids},
  volume = {237},
  pages = {105340},
  year = {2022},
  doi = {10.1016/j.compfluid.2022.105340},
}

@article{kalitzin2002,
  title = {Turbulence modeling in an immersed-boundary {RANS} method},
  author = {G. Kalitzin and G. Iaccarino},
  year = {2002},
  month = {01},
  pages = {415--426},
  journal = {Center for Turbulence Research Annual Research Briefs}
}

@article{kalitzin2005,
    author = {G. Kalitzin and G. Medic and G. Iaccarino and P. Durbin},
    title = {Near-wall behavior of RANS turbulence models and implications for wall functions},
    journal = {Journal of Computational Physics},
    month = {03},
    volume = {204},
    pages = {265--291},
    year = {2005},
    doi = {10.1016/j.jcp.2004.10.018},
}

@article{driver1985,
    author = {D. M. Driver and H. L. Seegmiller},
    title = {Features of a reattaching turbulent shear layer in divergent channel flow},
    journal = {AIAA Journal},
    volume = {23},
    number = {2},
    pages = {163--171},
    year = {1985},
    doi = {10.2514/3.8890},
}

@book{oF,
    author = {OpenCFD},
    title = {OpenFOAM: The Open Source CFD Toolbox. User Guide Version 1.4, OpenCFD Limited},
    year = {2007},
    publisher = {Reading UK},
}

@techreport{bredberg2000,
     author = {J. Bredberg},
     title = {On the wall boundary condition for turbulence models},
     group = {Department of Thermo and Fluid Dynamics},
     institution = {Chalmers University of Technology},
     number = {Internal report 00/4},
     address = {Goteborg},
     year = {2000},
}

@inproceedings{nasa,
    author = {C. Rumsey and B. Smith and G. Huang},
    title = {Description of a Website Resource for Turbulence Modeling Verification and Validation},
    booktitle = {40th Fluid Dynamics Conference and Exhibit},
    year = {2010},
    doi = {10.2514/6.2010-4742},
    url = {https://turbmodels.larc.nasa.gov/},
}

@book{moukalled2016,
    author = {F. Moukalled and M. Darwish and L. Mangani},
    title = {The finite volume method in computational fluid dynamics: an advanced introduction with {O}pen{FOAM} and {M}atlab},
    publisher = {Springer-Verlag},
    address = {Berlin, Germany},
    edition = {1},
    year = {2016},
    isbn = {978-3-319-16874-6},
}

@book{wilcox2006,
    author = {D.C. Wilcox},
    title = {Turbulence modeling for CFD},
    year = {2006},
    edition = {3},
    publisher = {DCW Industries, USA},
    isbn = {978-1-928729-08-2},
}

@article{constant2021,
    author = {B. Constant and S. Péron and H. Beaugendre},
    title = {An improved immersed boundary method for turbulent ﬂow simulations on Cartesian grids},
    journal = {Journal of Computational Physics},
    volume = {435},
    pages = {110240},
    year = {2021},
    doi = {10.1016/j.jcp.2024.113441},
}

@article{studenik2024,
    author = {O. Studeník and M. Isoz and M. Kotouč Šourek and P. Kočí},
    title = {Open{HFDIB}-{DEM}: {A}n extension to {O}pen{FOAM} for {CFD}-{DEM} simulations with arbitrary particle shapes},
    journal = {SoftwareX},
    volume = {27},
    pages = {101871},
    year = {2024},
    doi = {10.1016/j.softx.2024.101871},
}

@article{spalart2007,
    author = {P.R. Spalart and C.L. Rumsey},
    title = {Effective Inflow Conditions for Turbulence Models in Aerodynamic Calculations},
    journal = {AIAA Journal},
    volume = {45},
    number = {10},
    pages = {2544-02553},
    year = {2007},
    doi = {10.2514/1.29373},
}

@book{panton2013,
    author = {R.L. Panton},
    title = {Incompressible flow},
    year = {2013},
    edition = {1},
    publisher = {John Wiley \& Sons},
    isbn = {9781118013434},
}

@article{athkuri2023,
    author = {S.S.C. Athkuri and M.R. Nived and R. Aswin and V. Eswaran},
    title = {Computation of drag crisis of a circular cylinder using Hybrid {RANS}-{LES} and {URANS} models},
    journal = {Ocean Engineering},
    volume = {270},
    pages = {113645},
    year = {2023},
    issn = {0029-8018},
    doi = {10.1016/j.oceaneng.2023.113645},
}

@article{verzicco2023,
   author = {R. Verzicco},
   title = {Immersed Boundary Methods: Historical Perspective and Future Outlook},
   journal = {Annual Review of Fluid Mechanics},
   year = {2023},
   volume = {55},
   pages = {129--155},
   doi = {10.1146/annurev-fluid-120720-022129},
   publisher = {Annual Reviews},
}

@article{mittal2023,
    title = {Origin and evolution of immersed boundary methods in computational fluid dynamics},
    author = {Mittal, R. and Seo, J. H.},
    journal = {Physical Review Fluids},
    volume = {8},
    issue = {10},
    pages = {100501},
    numpages = {32},
    year = {2023},
    publisher = {American Physical Society},
    doi = {10.1103/PhysRevFluids.8.100501},
}

@article{cai2021,
    author = {S.-G. Cai and J. Degrigny and J.-F. Boussuge and P. Sagaut},
    title = {Coupling of turbulence wall models and immersed boundaries on Cartesian grids},
    journal = {Journal of Computational Physics},
    volume = {429},
    pages = {109995},
    year = {2021},
    doi = {10.1016/j.jcp.2020.109995},
}

@article{constant2024,
    author = {Benjamin Constant and Stéphanie Péron and Héloïse Beaugendre and Christophe Benoit},
    title = {An improved immersed boundary method for turbulent flow simulations on Cartesian grids: extension of a global geometric approach for thin boundary layers and strong flow incidence},
    journal = {Journal of Computational Physics},
    volume = {519},
    pages = {113441},
    year = {2024},
    doi = {10.1016/j.jcp.2024.113441},
}

@article{fadlun2000,
    author = {E.A. Fadlun and R. Verzicco and P. Orlandi and J. Mohd-Yusof},
    title = {Combined Immersed-Boundary Finite-Difference Methods for Three-Dimensional Complex Flow Simulations},
    journal = {Journal of Computational Physics},
    volume = {161},
    number = {1},
    pages = {35--60},
    year = {2000},
    doi = {10.1006/jcph.2000.6484},
}

@article{kim2001,
    author = {J. Kim and D. Kim and H. Choi},
    title = {An Immersed-Boundary Finite-Volume Method for Simulations of Flow in Complex Geometries},
    journal = {Journal of Computational Physics},
    volume = {171},
    number = {1},
    pages = {132--150},
    year = {2001},
    issn = {0021-9991},
    doi = {10.1006/jcph.2001.6778},
}

@article{uhlmann2005,
    author = {M. Uhlmann},
    title = {An immersed boundary method with direct forcing for the simulation of particulate flows},
    journal = {Journal of Computational Physics},
    volume = {209},
    number = {2},
    pages = {448--476},
    year = {2005},
    doi = {10.1016/j.jcp.2005.03.017},
}

@book{strang2014,
    author = {G. Strang},
    title = {Differential equations and linear algebra},
    year = {2014},
    publisher = {Wellesley-Cambridge Press},
    isbn = {978-0-980 2327-9-0}
}

@article{delassaux2021,
    title = {Sensitivity Analysis of Hybrid Methods for the Flow Around the Ahmed Body With Application to Passive Control With Rounded Edges},
    journal = {Computers \& Fluids},
    volume = {214},
    pages = {104757},
    year = {2021},
    doi = {10.1016/j.compfluid.2020.104757},
    author = {Francois Delassaux and Iraj Mortazavi and Emmanuelle Itam and Vincent Herbert and Charles Ribes},
}

@article{lundquist2010,
      author = {K.A. Lundquist and F. Katopodes Chow and J.K. Lundquist},
      title = {An Immersed Boundary Method for the Weather Research and Forecasting Model},
      journal = {Monthly Weather Review},
      year = {2010},
      publisher = {American Meteorological Society},
      address = {Boston MA, USA},
      volume = {138},
      number = {3},
      doi = {10.1175/2009MWR2990.1},
      pages = {796--817},
}

@article{das2018,
    author = {S. Das and A. Panda and N.G. Deen and J.A.M. Kuipers},
    title = {A sharp-interface Immersed Boundary Method to simulate convective and conjugate heat transfer through highly complex periodic porous structures},
    journal = {Chemical Engineering Science},
    volume = {191},
    pages = {1--18},
    year = {2018},
    doi = {10.1016/j.ces.2018.04.061},
}

@article{lavoie2022,
    author = {P. Lavoie and E. Radenac and G. Blanchard and E. Laurendeau and P. Villedieu},
    title = {Immersed Boundary Methodology for Multistep Ice Accretion Using a Level Set},
    journal = {Journal of Aircraft},
    volume = {59},
    number = {4},
    pages = {912--926},
    year = {2022},
    doi = {10.2514/1.C036492},
}

@article{lahooti2019,
    author = {M. Lahooti and D. Kim},
    title = {Multi-body interaction effect on the energy harvesting performance of a flapping hydrofoil},
    journal = {Renewable Energy},
    volume = {130},
    pages = {460--473},
    year = {2019},
    doi = {10.1016/j.renene.2018.06.054},
}

@article{blais2018,
    author = {B. Blais and F. Ilinca},
    title = {Development and validation of a stabilized immersed boundary CFD model for freezing and melting with natural convection},
    journal = {Computers \& Fluids},
    volume = {172},
    pages = {564--581},
    year = {2018},
    doi = {10.1016/j.compfluid.2018.03.037},
}

@article{tafti2022,
    author = {D.K. Tafti and L. He and K. Nagendra},
    title = {Large eddy simulation for predicting turbulent heat transfer in gas turbines},
    journal = {Philosophical Transactions of the Royal Society A: Mathematical, Physical and Engineering Sciences},
    volume = {372},
    number = {2022},
    pages = {20130322},
    year = {2014},
    month = {08},
    doi = {10.1098/rsta.2013.0322},
}

@article{saadat2018,
  author = {A. Saadat and C.J. Guido and G. Iaccarino and E.S.G. Shaqfeh},
  title = {Immersed-finite-element method for deformable particle suspensions in viscous and viscoelastic media},
  journal = {Physical Review E},
  volume = {98},
  issue = {6},
  pages = {063316},
  numpages = {17},
  year = {2018},
  month = {12},
  publisher = {American Physical Society},
  doi = {10.1103/PhysRevE.98.063316},
}

@article{breugem2012,
    author = {W.-P. Breugem},
    title = {A second-order accurate immersed boundary method for fully resolved simulations of particle-laden flows},
    journal = {Journal of Computational Physics},
    volume = {231},
    number = {13},
    pages = {4469--4498},
    year = {2012},
    doi = {10.1016/j.jcp.2012.02.026},
}

@article{lim2022,
    author = {S.H. Lim and S.G. Park},
    title = {Numerical analysis of energy harvesting system including an inclined inverted flag},
    journal = {Physics of Fluids},
    volume = {34},
    number = {1},
    pages = {013601},
    year = {2022},
    month = {01},
    doi = {10.1063/5.0072204},
}

@article{ouro2017,
    author = {P. Ouro and T. Stoesser},
    title = {An immersed boundary-based large-eddy simulation approach to predict the performance of vertical axis tidal turbines},
    journal = {Computers \& Fluids},
    volume = {152},
    pages = {74--87},
    year = {2017},
    doi = {10.1016/j.compfluid.2017.04.003},
}

@article{jenkins2016,
    author = {N. Jenkins and K. Maute},
    title = {An immersed boundary approach for shape and topology optimization of stationary fluid-structure interaction problems},
    journal = {Structural and Multidisciplinary Optimization},
    volume = {54},
    pages = {1191--1208},
    year = {2016},
    doi = {10.1007/s00158-016-1467-5},
}

@article{zhou2017,
    author = {C.H. Zhou},
    title = {{RANS} simulation of high-{R}e turbulent flows using an immersed boundary method in conjunction with wall modeling},
    journal = {Computers \& Fluids},
    volume = {143},
    pages = {73--89},
    year = {2017},
    doi = {10.1016/j.compfluid.2016.11.009},
}

@inproceedings{lienhart2003,
    author = {H. Lienhart and S. Becker},
    title = {Flow and Turbulence Structure in the Wake of a Simplified Car Model},
    booktitle = {SAE 2003 World Congress},
    month = {03},
    pages = {2003-01-0656},
    year = {2003},
    doi = {10.4271/2003-01-0656}
}

@inproceedings{drela1989,
    author = {M. Drela},
    editor = {J.T. Mueller},
    title = {{XFOIL}: An Analysis and Design System for Low Reynolds Number Airfoils},
    booktitle = {Low Reynolds Number Aerodynamics},
    year = {1989},
    publisher = {Springer Berlin Heidelberg},
    address = {Berlin, Heidelberg},
    pages = {1--12},
}

@inproceedings{ye2020,
    author = {H. Ye and Y. Chen and K. Maki},
    title = {A discrete-forcing immersed boundary method for turbulent-flow simulations},
    booktitle = {Proceedings of the Institution of Mechanical Engineers, Part M: Journal of Engineering for the Maritime Environment},
    month = {06},
    volume = {253},
    pages = {188--202},
    year = {2020},
    doi = {10.1177/1475090220927245}
}

@article{yu2025,
    author = {X. Yu and D.R. Fuhrman and Y. Shao},
    title = {Improved immersed boundary/wall modeling method for {RANS} solver coupled with wall functions: application to Cartesian grid systems},
    journal = {Engineering Applications of Computational Fluid Mechanics},
    volume = {19},
    pages = {2486657},
    number = {01},
    year = {2025},
    doi = {10.1080/19942060.2025.2486657},
}

@article{ong2009,
    author = {M.C. Ong and T. Utnes and L.E. Holmedal and D. Myrhaug and B. Pettersen},
    title = {Numerical simulation of flow around a smooth circular cylinder at very high Reynolds numbers},
    journal = {Marine Structures},
    volume = {22},
    pages = {142--153},
    year = {2009},
    doi = {10.1016/j.marstruc.2008.09.001},
}

@article{zhou2015,
    author = {B. Zhou and X. Wang and W.M. Gho and S.K. Tan},
    title = {Force and ﬂow characteristics of a circular cylinder with uniform surface roughness at subcritical Reynolds numbers},
    journal = {Applied Ocean Research},
    volume = {49},
    pages = {20--26},
    year = {2015},
    doi = {10.1016/j.apor.2014.06.002},
}

@article{onishi2021,
    author = {K. Onishi and M. Tsubokura},
    title = {Topology-free immersed boundary method for incompressible turbulence flows: {A}n aerodynamic simulation for “dirty” {CAD} geometry},
    journal = {Computer Methods in Applied Mechanics and Engineering},
    volume = {378},
    pages = {113734},
    year = {2021},
    doi = {10.1016/j.cma.2021.113734},
}

@incollection{murray2011,
    author = {M.T. Murray and M. Murray},
    title = {9 - High pressure die casting of aluminium and its alloys},
    editor = {R. Lumley},
    booktitle = {Fundamentals of Aluminium Metallurgy},
    publisher = {Woodhead Publishing},
    pages = {217--261},
    year = {2011},
    series = {Woodhead Publishing Series in Metals and Surface Engineering},
    isbn = {978-1-84569-654-2},
    doi = {10.1533/9780857090256.1.217},
}

@article{thompson1988,
    author = {B.E. Thompson and J.H. Whitelaw},
    title = {Flow-around airfoils with blunt, round, and sharp trailing edges},
    journal = {Journal of Aircraft},
    volume = {25},
    number = {4},
    pages = {334--342},
    year = {1988},
    doi = {10.2514/3.45568},
}

@misc{github2026,
    author = {L. Kubíčková and M. Isoz},
    title = {{openHFDIBRANS}},
    year = {2026},
    publisher = {GitHub},
    journal = {GitHub repository},
    howpublished = {\url{https://github.com/techMathGroup/openHFDIBRANS}}
}

@software{kubickova2026zenodo,
    author = {L. Kubíčková and M. Isoz},
    title = {{openHFDIBRANS}},
    month = {09},
    year = {2026},
    publisher = {Zenodo},
    doi = {10.5281/zenodo.22917049},
    url = {https://doi.org/10.5281/zenodo.22917049},
}

\begin{appendix}
\clearpage

\setcounter{figure}{0}
\setcounter{table}{0}
\section{Effective vs. perpendicular wall distance for complex geometries}
\label{app:effandperp}

\begin{figure}[htbp]
    \centering
    \includegraphics{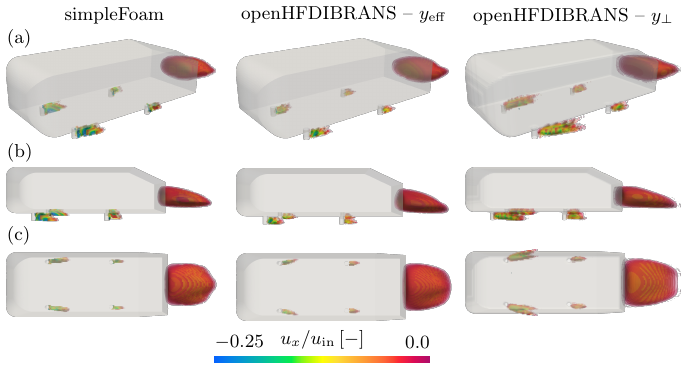}
    \caption{(a) Isometric, (b) side and (c) top view onto the recirculation regions behind the Ahmed body. Visualized as contours of the velocity field.}
    \label{fig:yorthorecirc}
\end{figure}
The Ahmed body benchmark, described in Section~\ref{sub:ahmed}, was used to compare cases when the effective distance $y_\mathrm{eff}$ or the perpendicular distance $y_\perp$ was used for the evaluation of $y^+_\mathrm{ib}$, see Section~\ref{sub:yplus}. The use of $y_\mathrm{eff}$ resulted in a significantly smaller recirculation region and improved agreement with simpleFoam, see Figure~\ref{fig:yorthorecirc}. Using $y_\perp$, the region length was $0.435\ \mathrm{m}$ and volume $0.0159\ \mathrm{m}^3$. In contrast, with $y_\mathrm{eff}$, the length and volume reduced to $0.356\ \mathrm{m}$ and $0.0094\ \mathrm{m}^3$, respectively. These values are in substantially closer agreement with the simpleFoam reference solution, which predicted a recirculation length of $0.377\ \mathrm{m}$ and volume of $0.0091\ \mathrm{m}^3$.

\setcounter{figure}{0}
\setcounter{table}{0}
\section{Sensitivity to numerical schemes}
\label{app:numschemes}

\begin{figure}[htbp]
    \centering
    \includegraphics{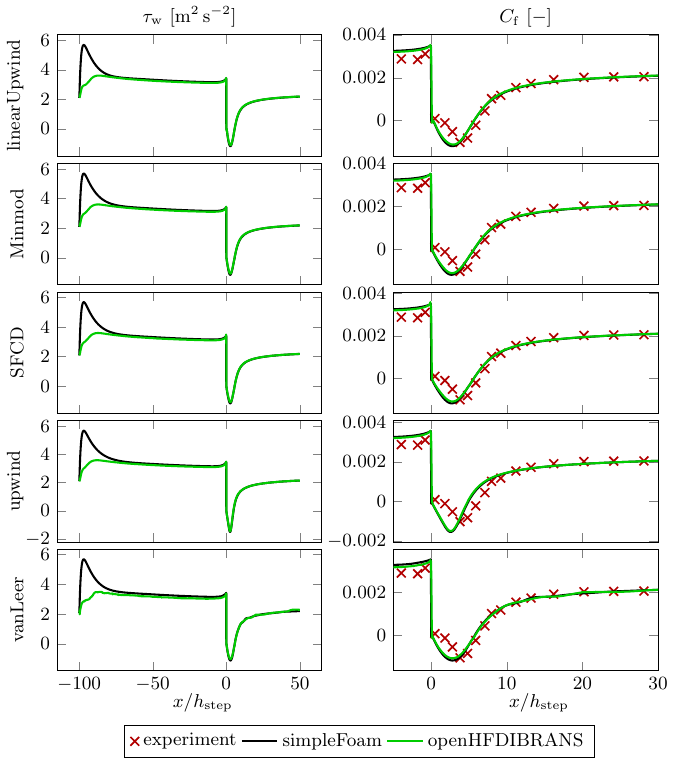}
    \caption{Evolution of the wall shear stress, $\tau_\mathrm{w}$, and skin friction coefficient, $C_\mathrm{f}$, in the backward facing step benchmark computed with different divergence schemes.}
    \label{fig:numschemesbfs}
\end{figure}

\begin{figure}[htbp]
    \centering
    \includegraphics{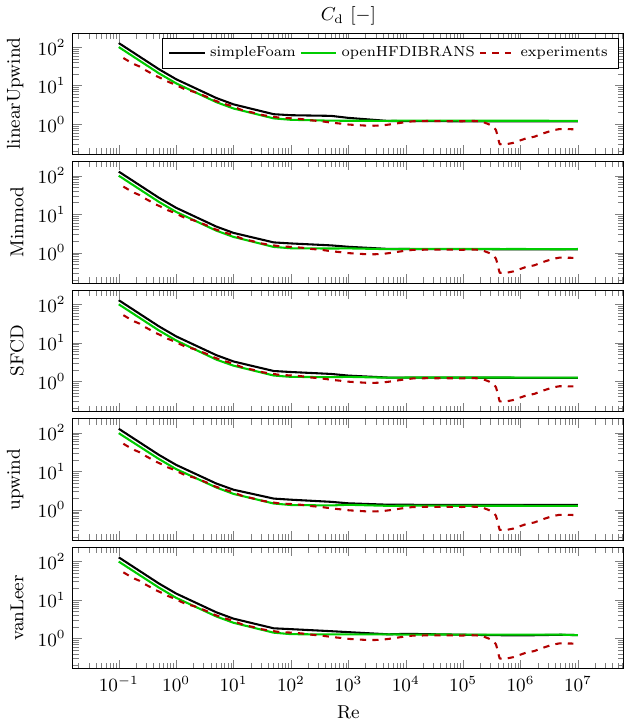}
    \caption{Evolution of the drag coefficient, $C_\mathrm{d}$, for a cylindrical obstacle in flow computed with different divergence schemes.}
    \label{fig:numschemescyl}
\end{figure}

\begin{figure}[htbp]
    \centering
    \includegraphics{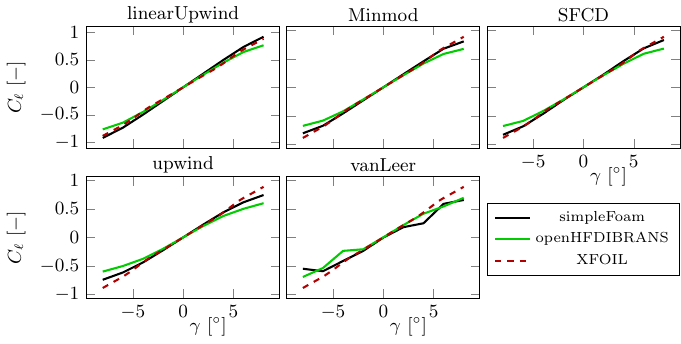}
    \caption{Evolution of the lift coefficient, $C_\mathrm{\ell}$, for for flow over a NACA-0009 airfoil computed with different divergence schemes.}
    \label{fig:numschemesnaca}
\end{figure}

The sensitivity of openHFDIBRANS to the choice of divergence scheme was comparable to that of simpleFoam. To assess this, three test cases were reevaluated:
\begin{inparaenum}[(i)]
    \item{the backward-facing step benchmark with the coarser mesh, Section~\ref{sec:bfs},}
    \item{flow over a cylindrical obstacle with mesh having $80$ thousand cells, Section~\ref{sub:cylinder}, and}
    \item{flow over a NACA-0009 airfoil with $\mathrm{Re} = 2\cdot10^6$, Section~\ref{sub:naca}.}
\end{inparaenum}

Comparison of the wall shear stress (backward-facing step), the drag coefficient evolution (cylinder), and lift coefficient evolution (NACA-0009) showed only minor differences between the discretization schemes, see Figures~\ref{fig:numschemesbfs}--\ref{fig:numschemesnaca}. For higher-order schemes, the results of both solvers shifted slightly towards the reference data, as expected. Moreover, the vanLeer scheme produced signs of instability in both solvers, most notably for the NACA-0009 case. Overall, the response of openHFDIBRANS to the choice of discretization scheme remained comparable to that of simpleFoam, indicating consistent numerical behavior.

\setcounter{figure}{0}
\setcounter{table}{0}
\section{Sensitivity to misalignment of mesh faces and immersed boundary}
\label{app:misalign}

\begin{figure}[htbp]
    \centering
    \includegraphics{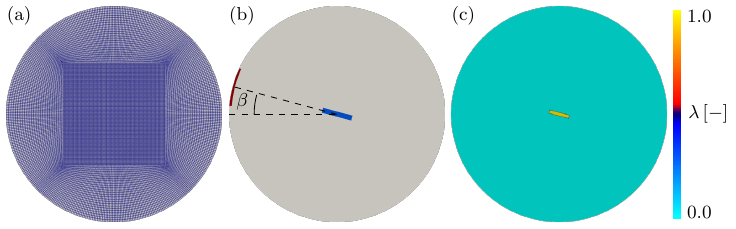}
    \caption{(a) Visualization of the mesh structure, the actual mesh contained $9\times$ more cells. Misalignment of (b) the inlet (red) and solid body (blue) positions and (c) lambda field by an angle $\beta = 15^\circ$.}
    \label{fig:misdoms}
\end{figure}
The sensitivity of openHFDIBRANS to misalignment between mesh faces and the immersed boundary was found to be low. To test the sensitivity, a circular computational domain with an orthogonal mesh in a central square region was considered, see Figure~\ref{fig:misdoms}a. A rectangular solid body was placed at the domain center and rotated by an angle $\beta$ to impose controlled misalignment with respect to the mesh faces, see Figure~\ref{fig:misdoms}b for an example with $\beta = 15^\circ$.

A portion of the outer boundary, indicated in red in Figure~\ref{fig:misdoms}b, was the inlet, while the rest of the boundary was the outlet. The inlet position was rotated consistently with the solid body to maintain a zero angle of attack with respect to the incoming flow. For $\beta = 15^\circ$, the resulting $\lambda$ field is shown in Figure~\ref{fig:misdoms}c.

\begin{figure}[htbp]
    \centering
    \includegraphics{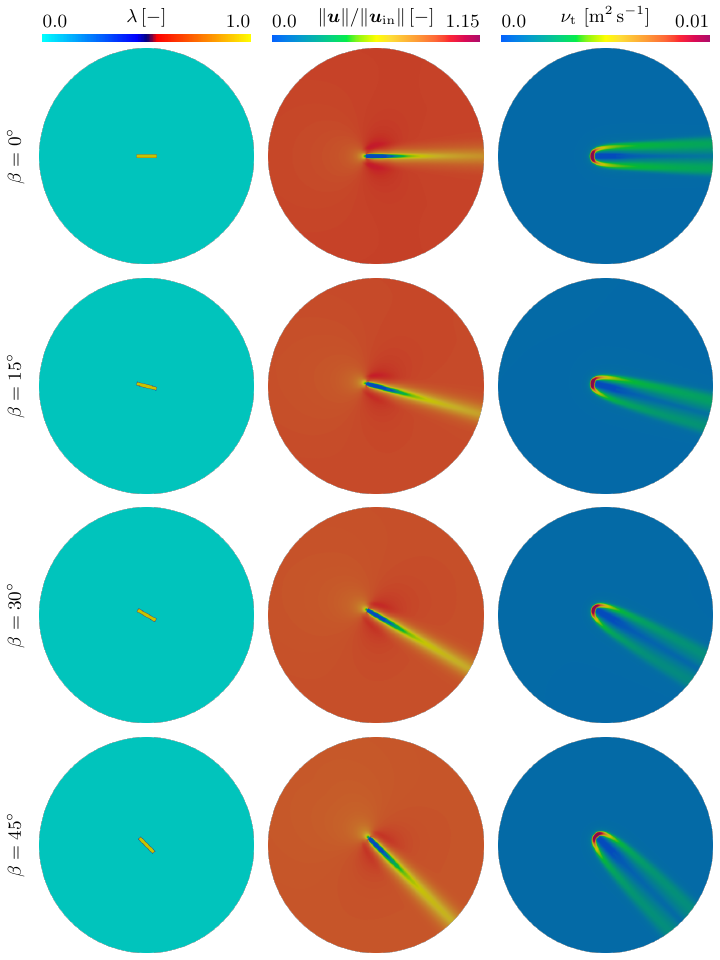}
    \caption{Comparison of the $\lambda$, $\|\bm{u}\|$ and $\nu_\mathrm{t}$ fields for varying misalignment angle $\beta$.}
    \label{fig:misfields}
\end{figure}
The boundary conditions were set in the same way as for the two-dimensional pipe flow test case described in Section~\ref{sec:straight}. The Reynolds number at the inlet was $\mathrm{Re} \sim 10^5$ and kinematic viscosity $\nu = 10^{-6}\ \mathrm{m}^2\,\mathrm{s}^{-1}$. The resulting $\|\bm{u}\|$ and $\nu_\mathrm{t}$ fields for angles $\beta \in (0,45)^\circ$, are depicted in Figure~\ref{fig:misfields}. For all misalignment angles, the flow features were preserved well, with only minor differences observed between the fields.

\setcounter{figure}{0}
\setcounter{table}{0}
\section{Effect of mesh-refinement}
\label{app:meshref}
While studying the effect of mesh-refinement on the simulation results, openHFDIBRANS converged towards the reference values at a rate comparable to simpleFoam. However, at coarse mesh resolutions, the openHFDIBRANS solver suffered from loss of geometric fidelity inherent to immersed boundary methods. The effect of mesh-refinement on the results was studied on the backward-facing step benchmark and flow around a NACA-0009 airfoil.

\begin{figure}[htbp]
    \centering
    \includegraphics{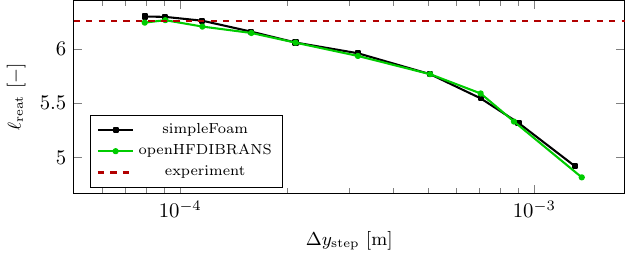}
    \caption{Evolution of the reattachment length $\ell_\mathrm{reat}$ in the backward-facing step benchmark for different mesh refinements defined by characteristic cell size in the step vicinity $\Delta y_\mathrm{step}$.}
    \label{fig:bfsmis}
\end{figure}
While changing the characteristic cell size in the backward-facing step benchmark, the reattachment length $\ell_\mathrm{reat}$ was monitored, and both openHFDIBRANS and simpleFoam showed very similar convergence behavior, see Figure~\ref{fig:bfsmis}. As the mesh coarsened, the reattachment length estimated by both solvers decreased.

\begin{figure}[htbp]
    \centering
    \includegraphics{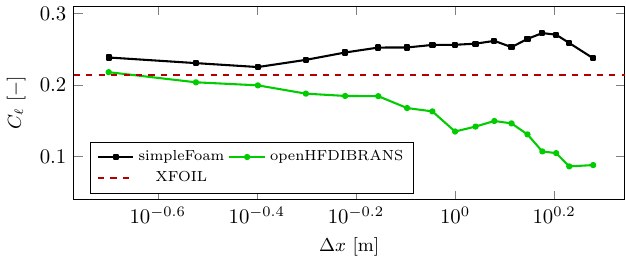}
    \caption{Evolution of the lift coefficient $C_\ell$ in simulation of flow over a NACA-0009 airfoil for different mesh refinements defined by a characteristic cell size $\Delta x$. The angle of attack was $\gamma = 2^\circ$ and $\mathrm{Re} = 2\cdot10^6$.}
    \label{fig:nacamis}
\end{figure}
For the NACA-0009 airfoil, the mesh-refinement study focused on the lift coefficient $C_\ell$ evaluated for $\gamma = 4^\circ$ at $\mathrm{Re} = 2\cdot10^6$. With finer meshes, both solvers converged to similar values, but when coarsening the mesh, the behavior of the solvers differed. While the simpleFoam estimate of $C_\ell$ gradually increased, the openHFDIBRANS one decreased. The likely reason is that with coarser meshes the geometry of the NACA-0009 becomes less resolved by the immersed boundary, with the most problematic part being the sharp tip of the trailing edge. On coarser meshes, the tip is represented by fewer cells causing effective rounding of its shape which affects the wake size~\citep{thompson1988}.

Overall, the mesh-refinement study indicated that openHFDIBRANS converges with a similar rate to simpleFoam. However, for very coarse meshes, the immersed boundary representation loses geometric fidelity more rapidly than the body-fitted approach.

\setcounter{figure}{0}
\setcounter{table}{0}
\section{Evolution of system matrix stiffness and solver convergence}
\label{app:stiffandconv}

\begin{figure}[htbp]
    \centering
    \includegraphics{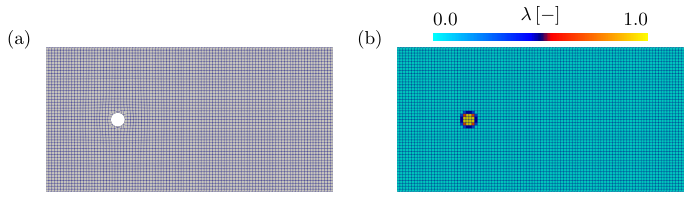}
    \caption{Computational domain and mesh used by (a) simpleFoam, (b) openHFDIBRANS colored by $\lambda$ field.}
    \label{fig:smallcyldoms}
\end{figure}
The convergence behavior and matrix conditioning of openHFDIBRANS were comparable to those of simpleFoam across all investigated Reynolds numbers, although openHFDIBRANS had slightly higher condition numbers for most system matrices. The solver convergence and matrix conditioning were assessed using a modified version of the simulation benchmark with a smooth cylinder, see Section~\ref{sub:cylinder}. To enable efficient evaluation of the discretization matrices eigenvalues, the computational mesh was significantly coarsened to $\sim\!5\cdot10^3$ cells, see Figure~\ref{fig:smallcyldoms}.

Simulations were performed for several Reynolds numbers and the $k$-$\omega$ turbulence model was used. For each of solved equations, the monitored quantities were:
\begin{inparaenum}[(i)]
    \item{the solution residual as calculated by OpenFOAM~\citep{oF}, and}
    \item{the condition number ($\kappa$) of the discretization matrix defined as}
\end{inparaenum}
\begin{equation}
    \kappa(M) = \frac{\max\left\{|\lambda^\mathrm{e}(M)|\right\}}{\min\left\{|\lambda^\mathrm{e}(M)|\right\}}
    \label{eq:kappa}
\end{equation}
where $\lambda^\mathrm{e}$ denotes the eigenvalues of the real square matrix $M$~\citep{strang2014}.

\begin{figure}[htbp]
    \centering
    \includegraphics{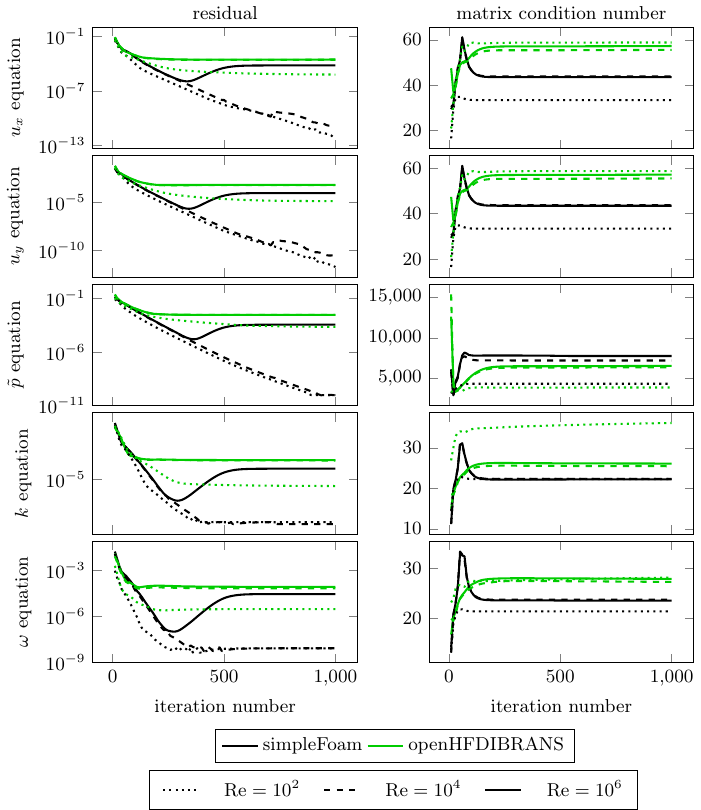}
    \caption{Evolution of the solution residual and matrix condition number with respect to the iteration number for all solved equations and three different $\mathrm{Re}$.}
    \label{fig:resandcon}
\end{figure}
In terms of convergence, simpleFoam outperformed openHFDIBRANS, particularly for lower Reynolds numbers, see left column in Figure~\ref{fig:resandcon}. For the highest Reynolds number case, however, the residual evolution was nearly identical for both solvers. The superiority of the body-fitted approach is to be expected. Nevertheless, owing to the $\lambda$-masked \revs{fluxes in} the continuity equation, \revs{$\nabla \cdot \left[\left(1 - \lambda\right)\bm{u}\right] = 0$}, openHFDIBRANS did not exhibit pressure oscillations near the immersed fluid-solid interface and the residual evolution remained relatively smooth for all variables and tested $\mathrm{Re}$.

Next, the condition numbers $\kappa$ of matrices originating from simpleFoam and openHFDIBRANS were of the same order of magnitude, see right column in Figure~\ref{fig:resandcon}. However, for velocity and turbulence variables, the matrices from simpleFoam were consistently better conditioned than those obtained with openHFDIBRANS. The pressure equation was the most ill-conditioned for both solvers. Apart from an initial spike observed in openHFDIBRANS, the evolution of $\kappa$ during the solver iterations was remarkably similar across all tested Reynolds numbers. This initial spike can be attributed to the $\lambda$-masked continuity equation formulation, which directly modifies the structure of the pressure equation.

\begin{figure}[htbp]
    \centering
    \includegraphics{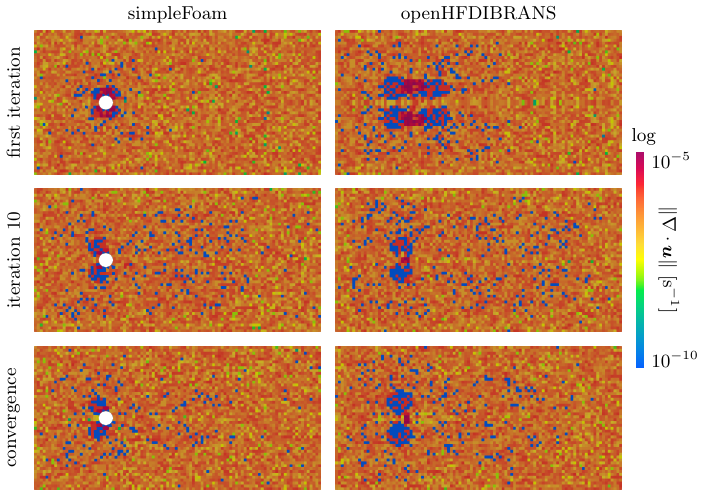}
    \caption{Comparison of the local continuity errors quantified by the magnitude of the velocity divergence for flow over a smooth cylinder with $\mathrm{Re} = 10^6$.}
    \label{fig:divphi}
\end{figure}
A key limitation of the $\lambda$-mask\revs{ing in the} continuity equation is the lack of explicit enforcement of mass conservation over the entire computational domain. During the first solver iterations, the immersed boundary produced localized source and sink regions, see Figure~\ref{fig:divphi}. However, upon convergence, the level of mass conservation in openHFDIBRANS became comparable to that of simpleFoam. Mass conservation across the immersed boundary stems from the fact that the $\lambda$-masked continuity equation still enforces continuity in a weak sense across the domain. In particular, the resolved mass balance in fluid cells adjacent to the immersed boundary suppresses spurious fluxes, thereby ensuring consistent mass conservation across the interface.

\setcounter{figure}{0}
\setcounter{table}{0}
\section{Parallel scaling}
\label{app:parallel}
The three-dimensional Ahmed body benchmark was used to examine the parallel scaling of openHFDIBRANS. No major differences from simpleFoam are expected, since the same matrix solvers are utilized. The differences are confined to modifications of the governing equations and a small number of additional operations within the solver loop that are specific to openHFDIBRANS.

\begin{figure}[htbp]
    \centering
    \includegraphics{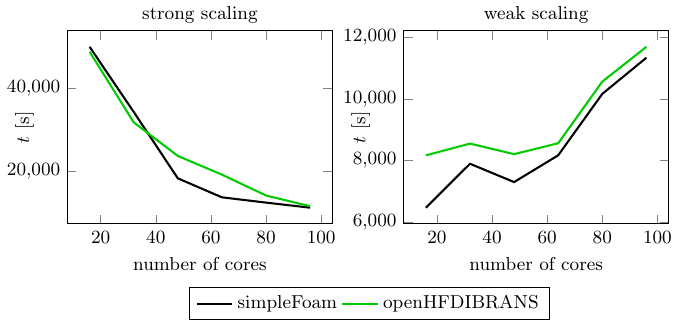}
    \caption{Comparison of simulation times with respect to the number of cores used while having fixed the total number of cells (strong scaling) and the number of cells per core (weak scaling).}
    \label{fig:parscaling}
\end{figure}
The Ahmed body test case as presented in Section~\ref{sub:ahmed} had a mesh with $\sim2\cdot10^7$ cells and was initially run in parallel on $96$ cores, corresponding to roughly $2\cdot10^5$ cells per core. Both strong and weak scaling were considered. Strong scaling was evaluated by rerunning the simulation on the original mesh while reducing the number of cores. For weak scaling, reduced-size meshes were generated and executed on appropriately chosen core counts to maintain a constant number of cells per core. Simulation times for both scaling scenarios showed no significant differences between openHFDIBRANS and simpleFoam, see Figure~\ref{fig:parscaling}. This outcome is consistent with expectations and indicates that the here-presented immersed-boundary formulation does not introduce a noticeable overhead in parallel performance.
\end{appendix}

\end{document}